\documentclass[bibyear]{latex_style/aa} 
\usepackage{latex_style/natbib}

\usepackage[section]{algorithm}

\usepackage{algorithmic}
\usepackage{amsmath}
\usepackage{caption} 
\usepackage{color}
\usepackage{graphicx}
\usepackage{hyperref}
\usepackage{mathcomp}
\usepackage{mathtools}
\usepackage{mdwmath}
\usepackage{mdwtab} 
\usepackage{pifont} 
\usepackage{subfig}
\usepackage{txfonts}
\usepackage{url}

\RequirePackage{color,graphicx}

\definecolor{blue}{rgb}{0,0.5,1}
\definecolor{bred}{rgb}{0.64,0,0}
\definecolor{green}{rgb}{0,0.5,0}
\definecolor{linkcolour}{rgb}{0.64,0,0}

\def\simlt{\lower.5ex\hbox{$\; \buildrel < \over \sim \;$}}
\def\simgt{\lower.5ex\hbox{$\; \buildrel > \over \sim \;$}}

\hypersetup{colorlinks,breaklinks,urlcolor=linkcolour, linkcolor=linkcolour}

\bibpunct{(}{)}{;}{a}{}{,}

\begin{document}
	
	\title{
Accelerating the cosmic microwave background map-making procedure through preconditioning}

	\titlerunning{Accelerating the CMB map-making through preconditioning}
	
	\author{
    	M. Szydlarski\inst{1}\thanks{\emph{Present address:} Institute of Theoretical Astrophysics, University of Oslo, PO Box 1029, Blindern, 0315 Oslo, Norway. New \email{mikolaj.szydlarski@astro.uio.no}}
	        \and
		L. Grigori\inst{2}
		\and
		R. Stompor\inst{3}
	}
	
	\institute{
                INRIA Saclay-\^Ile de France, F-91893 Orsay, France,
		\email{mikolaj.szydlarski@inria.fr} 
       \and
                 INRIA Rocquencourt, Alpines, B.P. 105, F-78153 Le Chesnay Cedex
                 and UPMC Univ Paris 06, CNRS UMR 7598, Laboratoire Jacques-Louis Lions, F-75005, Paris, France,
		\email{laura.grigori@inria.fr} 
	\and
		AstroParticule et Cosmologie, Univ Paris Diderot, CNRS/IN2P3, CEA/Irfu, Obs de Paris, Sorbonne Paris Cit\'e, France\\
		\email{radek@apc.univ-paris-diderot.fr}
	}

   \date{Received; accepted}


\abstract{ Estimation of the sky signal from sequences of time ordered data
  is one of the key steps in cosmic microwave background (CMB)
  data analysis, commonly referred to as the map-making problem. Some
  of the most popular and general methods proposed for this problem
  involve solving generalised least squares (GLS) equations with
  non-diagonal noise weights given by a block-diagonal matrix with
  Toeplitz blocks.
    In this work, we study new map-making solvers potentially suitable
  for applications to the largest anticipated data
  sets. They are based on iterative conjugate
  gradient (CG) approaches enhanced with novel, parallel, two-level
  preconditioners.  
We apply the proposed solvers
  to examples of simulated non-polarised and polarised CMB
  observations and a set of idealised scanning strategies with sky
  coverage ranging from a nearly full sky down to small sky
  patches.
  We discuss their implementation
  for massively parallel computational platforms and their performance
  for a broad range of parameters that characterise the simulated data sets in detail.
We find that our best new solver can outperform carefully optimised standard solvers used today by 
a factor of as much as five in terms
  of  the convergence rate and a factor of up to four in terms of the time to solution, without
  significantly increasing the memory consumption and the volume of
  inter-processor communication. The performance of the new algorithms
  is also found to be more stable and robust and less dependent on
  specific characteristics of the analysed data set. We therefore
  conclude that the proposed approaches are well suited to address successfully 
  challenges posed by new and forthcoming CMB data sets.
}



\maketitle 



\section{Introduction}\label{sec:introduction} 

The majority of current and anticipated Cosmic Microwave Background (CMB) experiments scan the sky with 
large arrays of detectors, producing as the result time-ordered data, composed of many billions of samples.
One of the key steps of the CMB data analysis consists 
of recovery of the sky signal from these huge noisy, time-ordered
data set. This can be phrased as a linear inverse
problem if some of the crucial characteristics of the
instrument and its operations are known.  These typically include
some knowledge of the noise properties of all detectors, an instrumental 
response function, or an observation direction at each measurement.
A well-known solution to such a problem
is given in a form of a generalised least squares (GLS) equation with
weights given by an arbitrary, symmetric positive definite
matrix~\citep[][]{Tegmark97a}. If the weights are taken to be equal to the inverse
covariance of the time-domain noise, this estimate  is also  
a minimum variance and a maximum likelihood solution to the problem.  Its computation typically requires
either an explicit factorisation of a huge
matrix \citep[][]{Tegmark97b, Borrill99, Stompor02} or some iterative
procedure \citep[][]{Wright96, Oh99, Dore01}, involving multiple, large matrix-vector products at each iteration.  The map-making problem is
therefore primarily numerical and is a particularly challenging problem given sizes of the current and forthcoming CMB data sets. These not only
determine the number of floating point operations (flops), which need to be
performed  to calculate the solution, but also sizes of the arrays, which have to be manipulated on
and stored in the computer memory. These set requirements on the computer resources, which often can be
only matched by massively parallel computing platforms. The GLS approach
indeed has been frequently used by many past and current modern CMB data analysis efforts, requiring parallel numerical algorithms 
and their efficient implementation.

As of today, essentially all existing map-making codes implementing
the GLS solution resort to the
same iterative technique based on a preconditioned conjugate gradient
(PCG) method with a preconditioner that corresponds to the system matrix of the GLS
problem computed under a hypothesis of white noise in the time
domain~\citep[however, see, e.g.,][for some exceptions restricted to special scanning strategies.]{WandeltHansen2003,  NaesLouis2013}. 
 This approach has been extensively used in the analysis of
diverse CMB data sets and found to be very efficient and suitable for parallelisation. However, its current implementations are
 unlikely to meet the demands of future data sets, even if
the projected increase in computational power is accounted for.
This performance gap can be addressed by either further optimising of the numerical
implementations of the current approach, or by devising better alternative algorithms.

In the spirit of the latter, we recently proposed in ~\citet{midas_sc12} a
new preconditioner suitable for the GLS map-making problem and studied
its properties in the context of
simulated CMB-like, total intensity data sets.  The aim of this paper
is to further generalise this technique, extending
it to accommodate polarisation-sensitive data sets and to evaluate
its performance on a set of realistic simulations of CMB
observations. We also discuss the role of some of the 
fundamental map-making parameters, such as
the noise correlation length, on the performance of the standard and
new map-making algorithms and the quality of the derived estimates.



\section{Problem definition}\label{sec:problem_def}

A measurement, $\vec{d_{t}}$, performed at time $t$ by a detector of a
typical scanning CMB instrument can be modelled in general as a sum of two
contributions, a sky signal, $\vec{s}_{t}$, and noise of the
instrument, $\vec{\breve{n}_{t}}$. The sky signal, $\vec{s}_t$,
corresponds to the signal in the direction in which the instrument was
pointing to at time $t$. Treating the sky signal as pixelized and
denoting the number of measurements by ${\cal N}_{t}$ and 
 the number of sky pixels by ${\cal
  N}_{p}\;(\ll {\cal N}_{t})$, the measured
signal can be represented as the result of the application of a
(${\cal N}_{t} \times {\cal N}_{p}$) projection operator,
$\mathbf{P}_{tp}$, to the sky map, $\vec{x_{p}}$.  The data model can
be consequently written as,
\begin{eqnarray}\label{eq:dataModelGen}
	\vec{d_t} = \vec{s}_{t} + \vec{\breve{n}_{t}} = \mathbf{P}_{tp} \, \vec{x_{p}} + \vec{\breve{n}_{t}}.
\end{eqnarray}
Hereafter, we refer to the projection operator, $\mathbf{P}_{tp}$, 
as the pointing matrix, and to  vectors $\vec{d}_t$ and $\vec{\breve{n}}$ 
as the time domain data and noise streams, respectively.

In general, the structure of the pointing matrix can be
quite complex, as it may represent an instrumental beam
convolution~\citep[e.g.,][]{ArmitageWandelt2004, Harrison_etal_2011,
  KeihanenRainecke2012}. However, if, for simplicity, we assume that
the instrumental beams are axially symmetric and consider the
sought-after sky maps, $\vec{{x}_p}$, as already convolved with the
beam, the resulting pointing matrices are typically very
sparse. For example, for a total intensity measurement obtained by a single
dish experiment, the pointing matrix has only one non-zero entry
per row corresponding to the pixel, $p$, observed at time $t$. For a
polarisation-sensitive, single dish experiment, the map,
$\vec{x_p}$, is composed of three sky maps corresponding to three
potentially non-zero Stokes parameters, I, Q, and U, which are used to describe 
 the linear polarised CMB radiation. The signal
detected by the experiment is a linear combination of three
Stokes amplitudes measured in the same pixel on the sky at a given time. Consequently, the
pointing matrix has three non-zero entries per row, 
corresponding to three Stokes parameters. If we
observe pixel $p$ at time $t$ and denote the values of the Stokes
parameters in this pixel as $\vec{i}_p, \vec{q}_p$, and $\vec{u}_p$,
we can then write
\begin{eqnarray}\label{eq:signal_with_polarization}
	\vec{s_{t}} & = &  \vec{i}_{p} + \vec{q}_{p} \cos 2\,\phi_{t} + \vec{u}_{p} \sin 2\,\phi_{t},
\end{eqnarray}
where $\vec{\phi}_t$ is the orientation of the polariser with respect to the sky coordinates at time $t$. The elements of the first vector on the right hand side of this equation define the only three non-zero values in the row
of the pointing matrix that correspond to time $t$. For a total intensity measurement, the second and third of these elements are also zero, and its Q and U entries can be removed from the solution vector as their value can not be recovered from the data. Consequently, the size of the estimated map is smaller, and the pointing matrix has only one non-zero per row as mentioned earlier. In this way, the latter case can be looked as a subcase of the polarised one and we focus on the former in the following.
The structure of the pointing matrix therefore reflects the type of the experiment and also its scanning strategy, defining which sky pixel is observed at time $t$.
Hereafter, we assume that the pointing matrix is full column-rank as we restrict ourselves solely to the sky pixels, which are actually observed. We refer to the operation $\mathbf{P}\,\vec{x}$, where $\vec{x}$ is a 
map-like vector, as the de-pointing operation, to the transpose operation, $\mathbf{P}^t\,\vec{y}$, where $\vec{y}$ is a time domain vector, as the pointing operation.

\subsection{Generalized least squares problem}\label{sub:the_maximum_likelihood_solution} 

A generalised least squares (GLS) solution to the problem in
equation~(\ref{eq:dataModelGen}) is given
by~\citep[e.g.,][]{bj1996numerical},
\begin{eqnarray}
\vec{m} = \left(\mathbf{P}^{\,t}\,\mathbf{M}\,\mathbf{P}\right)^{-1}\,\mathbf{P}^{\,t}\,\mathbf{M}\,\vec{d},
\label{eq:glsSol}
\end{eqnarray}
where $\mathbf{M}$ is a non-negative definite symmetric matrix and
$\vec{m}$ is an estimate of the true sky signal, $\vec{s}$.  If the
instrumental noise, $\vec{\breve{n}}$, is Gaussian and characterised
by covariance $\mathbf{N}$, then the maximum likelihood (as well as
minimum) variance estimates of the sky signal are given by
equation~(\ref{eq:glsSol}), if $\mathbf{M}$ is selected to be
equal to the inverse noise matrix, $\mathbf{M} =
\mathbf{N}^{-1}$. We note that whatever the choice of this matrix,
the GLS solution results in an unbiased estimate of the true sky
signal, at least as long as the pointing matrix is
correct. Nevertheless, to obtain the map estimate of
sufficient quality, it is usually important to ensure that $\mathbf{M}$ is 
close to the inverse noise covariance.

The instrumental noise is typically well described
as a Gaussian piece-wise stationary process. Its covariance
$\mathbf{N}=\langle\vec{\breve{n}}\,\vec{\breve{n}}^{T}\rangle$ is
therefore a block-diagonal matrix with Toeplitz blocks 
corresponding to different stationary time pieces. Though the inverse
of a Toeplitz matrix is not Toeplitz, it has been
suggested~\cite[][]{Tegmark97b} and shown~\citep[e.g.,][]{Stompor02}
that precision sufficient for most practical purpose, that $\mathbf{N}^{-1}$ can be approximated as a block-diagonal matrix with
Toeplitz blocks, which are constructed directly from the corresponding
blocks of the covariance itself.  We therefore assume throughout this
work that
\begin{eqnarray}
\mathbf{M} = \mathbf{N}^{-1}\simeq
\begin{pmatrix}\label{eq:N_def}
\mathbf{T}_{0} & 0 & \dots & 0\\
0 & \mathbf{T}_{1} &  \dots & 0\\
\vdots &  & \ddots & 0\\
0 & \dots &\dots & \mathbf{T}_{K}
\end{pmatrix},
\end{eqnarray}
where $\mathbf{T}_{i}$ denotes a symmetric Toeplitz block of a size $t_i
\times t_i$, and therefore $\sum_{i=0}^{K}\,t_i \, = \, {\cal
  N}_t$. The noise covariance (as well as its inverse) is a matrix of
size ${\cal N}_t \times {\cal N}_t$, where $N_t \ga {\cal O}(10^9)$ for modern observations.  
However, if
the structure of the matrix is explicitly accounted for, the memory required for its storage is
of the same order as the memory needed to store the data vector,
$\vec{d}$.
Another advantage of Toeplitz matrices is that they can be efficiently multiplied by a vector using Fast
Fourier Transforms (FFTs)~\citep[e.g.,][]{golub1996}.

We note that further approximations of the inverse noise
covariance are possible and often employed. For instance, the
Toep\-litz blocks can be assumed to be band-diagonal
~\citep[e.g.,][]{Stompor02}. We discuss this  option in detail later on.

The problem of reconstructing the underlying sky signals 
therefore amounts to efficiently solving equation~\eqref{eq:glsSol} given the
assumptions about the noise covariance listed previously. Given
the sizes of time-domain data sets of current and forthcoming
CMB experiments, this has been shown to be rather challenging, requiring
advanced parallel numerical algorithms.

A direct computation of such a solution requires an explicit
construction and inversion of the GLS system matrix,
\begin{eqnarray}\label{eq:Sdef}
\mathbf{A} & \equiv & \mathbf{P}^{\,t}\mathbf{N}^{-1}\mathbf{P}.
\end{eqnarray}
This matrix has a size of ${\cal N}_p \times {\cal N}_p$ and is in general dense and without any universal structure.
Consequently, both of these operations require ${\cal O}({\cal N}_p^3)$ flops.
What turned out to be sufficient for some of the first big CMB data sets~\citep[][]{boom, maxima} is thus prohibitive for the data sizes considered here.

Alternately, the problem  can be rephrased as a linear system~\citep[e.g.][]{Oh99, madmap},
\begin{eqnarray}\label{eq:linear_system}
	\mathbf{A}\, \vec{x}   & = & \vec{b},
\end{eqnarray}
with  matrix $\mathbf{A}$  as given above and 
the right-hand side given by
\begin{eqnarray}
	\vec{b}      & \equiv & \mathbf{P}^{\,t}\,\mathbf{N}^{-1} \, \vec{d}, \label{eq:rhs}
\end{eqnarray}
and its solution obtained with help of some iterative linear system solver~\citep[e.g.][]{golub1996}. 


Iterative solvers repeatedly perform matrix-vector products of the system matrix, $\mathbf{A}$, and a vector. This is the case for the so-called 
Krylov space methods, which are the methods considered in this work (see Appendix~\ref{app:aprox_eigenvector} for some background). The question
therefore arises whether the matrix-vector product can be accomplished efficiently in our case and in particular without ever explicitly constructing the matrix.  
This indeed turns out to be possible by representing the matrix, 
$\mathbf{A}$,  as in equation~\eqref{eq:Sdef}, and performing
the required matrix-vector operations from right to
left~\citep[see, for example,][]{madmap}. That is,
\begin{eqnarray}\label{eq:a_times_x}
	\mathbf{A}\,\mathbf{x} \, \equiv\, \mathbf{P}^{\,t}\Big[\mathbf{N}^{-1}\big[\mathbf{P}\vec{x}\big]\Big].
\end{eqnarray}
This exchanges the product of the dense, unstructured matrix, $\mathbf{A}$, for a sequence of operations starting with
 a depointing operation, $\mathbf{P}\mathbf{x}$, followed by a noise weighting, $\mathbf{N}^{-1}\mathbf{y}$, and a pointing 
 operation, $\mathbf{P}^t\mathbf{y}$. Here $\mathbf{x}$ and $\mathbf{y}$ are arbitrary pixel and time domain vectors.
As all these operations involve either highly structured, $\mathbf{N}^{-1}$, or sparse, $\mathbf{P}$, operators, one can expect that 
they can be performed very efficiently, as has been found to be the case~\citep[e.g.,][]{madmap}.

The computational complexity of the map-making algorithm described
above is ${\cal O}({\cal N}_{iter} \times {\cal N}_{t}\,(1 + \sum_i
\log_{2}\, t_{i}))$, where ${\cal N}_{iter}$ represents the
number of iterations of the iterative solver with the linear term in the data size, ${\cal N}_{t}$,
quantifying the cost of the pointing and depointing operations and the logarithmic term the cost of fast Fourier Transforms. In the latter case,  the block-diagonal structure of $\mathbf{N}^{-1}$ has been
explicitly accounted.  One obvious way
to shorten the computational time is to reduce the number of
iterations, ${\cal N}_{iter}$, required to reach a desired precision.
Another way is to try to cut the time needed to perform each
step. This could be achieved either by employing better algorithms or
introducing some additional assumptions.  


There are currently several implementations of iterative solvers developed in the CMB map-making context, such as MapCUMBA~\citep{Dore01},
ROMA~\citep{ROMA2005}, SANEPIC~\citep{sanepic},
MINRES~\citep{minres}, and MADmap~\citep{madmap}.  
They all employ a preconditioned conjugate gradient (PCG) method and, as defined at the beginning of the next section, seek efficiency in better implementations of the operations perfomed in equation~\eqref{eq:a_times_x} while adhering to the same set of the most straightforward preconditioners.

In this work, we instead focus on the iterative solvers themselves
 and propose new, more efficient
preconditioning techniques suitable for the map-making
problem. We then compare their performance with that of the standard technique, discussing approaches to improving the performance of the latter in this context.



\section{Preconditioned iterative solvers}\label{sec:iterative_solvers}

\subsection{Preliminaries and motivation}\label{sec:prec_prelim}

It is well known~\citep[see][]{CGfirst,sluis86} that the convergence
rate of the conjugate gradient method applied to solving a linear
system with a symmetric positive definite (SPD) system  matrix,
as in equation~\eqref{eq:linear_system}, depends on the
distribution of the eigenvalues of the system matrix.  Indeed,
it can be shown that we have
\begin{eqnarray}\label{eq:err_bound}
	\|{\vec{x}}-{\vec{x}}_{j}\|_{A} \leq 2 \| {\vec{x}} - {\vec{x}}_{0} \|_{A}\left(\frac{\sqrt{\kappa}-1}{\sqrt{\kappa}+1}\right)^{j},
\end{eqnarray}
after $j$ iterations of CG, where ${\vec{x}}_{0}$ is an initial guess for the solution $\vec{x}$ and $\|\vec{x}\|_{A}$ is a norm of $\vec{x}$ defined as
$\|\vec{x}\|_{A}=\sqrt{\vec{x}^{T}\mathbf{A}\vec{x}}$. The condition
number, $\kappa = \kappa(\mathbf{A})$, is given by the ratio of the
largest to the smallest eigenvalue of $\mathbf{A}$.  To accelerate the
convergence of CG, one may therefore opt to solve a preconditioned
system $\mathbf{M}\,\mathbf{A}{\vec{x}} = \mathbf{M}\vec{b}$, where
the preconditioner $\mathbf{M}$ is chosen so that
$\mathbf{M}\,\mathbf{A}$ has a smaller condition number than
$\mathbf{A}$ and/or a more clustered eigenspectrum.

To date nearly all studies of the PCG solver in context of the
map-making problem  have relied on the same, intuitive, easy to compute
and implement preconditioner~\citep[e.g.,][]{madmap}, defined as,
\begin{eqnarray}\label{eq:stdPrec}
 \mathbf{M}_{BD} =  \left(\mathbf{P}^{T} \, {\rm diag}\, (\mathbf{N}^{-1}) \, \mathbf{P}\right)^{-1}.
\end{eqnarray}
Here, ${\rm diag}\, (\mathbf{N}^{-1}) $ is a diagonal matrix consisting
of the diagonal elements of $\mathbf{N}^{-1}$. $\mathbf{M}_{BD}$ is
the inverse of $\mathbf{A}$ whenever the time-domain
noise is diagonal, that is ${\rm diag}\, (\mathbf{N}^{-1}) =
\mathbf{N}^{-1}$.  It can therefore be expected to provide a
good approximation to the inverse of the system matrix, $\mathbf{A}^{-1}$, in more general cases
and hence to be efficient as a preconditioner. Given the
assumptions made here about the pointing matrix the
preconditioner, $\mathbf{M}_{BD}$, is a block diagonal, and the sizes
of the blocks are equal to the number of Stokes
parameters. We will therefore refer to it in the following as either
the standard or the block-diagonal preconditioner.

\begin{figure}[!t]
\centering
\includegraphics[width=0.47\textwidth]{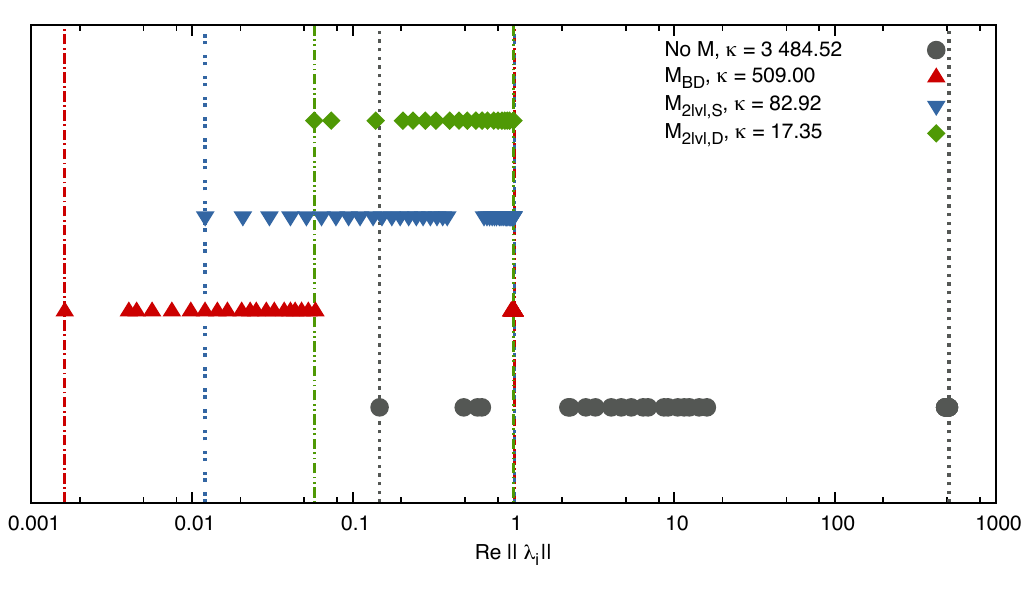}
\caption{An approximated spectrum (twenty of the smallest and largest eigenvalues) of an example system matrix derived
for the polarisation-sensitive case and
  preconditioned with different preconditioners. From the bottom to the top, we show the eigenvalues of
  $\mathbf{A}$,  $\mathbf{M}_{BD}\,\mathbf{A}$, and $\mathbf{M}_{2lvl,\{S,D\}}\,\mathbf{A}$, where $\mathbf{M}_{2lvl,\{S,D\}}$ denotes
  the two-level preconditioners proposed in Section~\ref{sec:two-level-prec}. The respective condition numbers, $\kappa$, are  given in the legend.
  }
  
\label{fig:eigen_values}
\end{figure}

The impact of this preconditioner on the eigenspectrum of the system
matrix is illustrated in Figure~\ref{fig:eigen_values},
where we compare the eigenspectra of the preconditioned and actual system matrix. It
is clear from the figure that the preconditioner performs very well as far
as very large eigenvalues of the initial matrix are concerned as it
shifts them to nearly unity.  This leads to an overall improvement
of the system matrix condition number, as large and small eigenvalues
are clearly rescaled differently. Nevertheless, small eigenvalues persist and can potentially
continue to hinder the convergence of the iterative solvers. However,
the number of these small eigenvalues seems limited in the cases of interest here. Indeed, in the example 
shown in the Figure with red triangles, there are fewer than $20$ eigenvalues smaller than a factor of $\sim 20$ than
the largest one, which, however, span a range of nearly two orders of magnitude.
This suggests that we could significantly improve on the condition number
of the preconditioned system matrix, if we could scale up these small eigenvalues
and do so without affecting significantly the others. This observation leads to
the idea of a two-level preconditioner discussed in the next Section.

\subsection{Two-level preconditioners} \label{sec:two-level-prec}

Let us first consider a numerically efficient preconditioner,
  which can selectively rescale a set of the smallest eigenvalues of
  the system matrix.  This can be done with a technique called
  deflation.  Deflation has been an active research area in numerical
  algebra, since the late eighties (see, e.g.,~\citet{Nicolaides:1987}
  or the recent survey
  \citet{gutknecht12:_spect_deflat_in_krylov_solver} and references
  therein) and successfully employed in diverse scientific and
  industrial contexts.  In essence, it aims at deflating the matrix
  from an unwanted subspace which hinders the convergence of iterative
  methods.
This unwanted subspace is typically the subspace spanned by
eigenvectors corresponding to the eigenvalues close to zero of the
system matrix~\citep{Morgan_GMRES,
  kharchenko95:_eigen_trans_based_precon_for, Chapman96deflatedand}.
We use a deflation method based on the deflation matrix
$\mathbf{R}$\citep[][]{TANG:2009} in our work defined as 
\begin{eqnarray}\label{eq:projection}
\mathbf{R} := \mathbf{I} -
\mathbf{A}\,\mathbf{Z}\, \mathbf{E}^{-1}\,\mathbf{Z}^{T}\, ,\quad
\mathbf{E} := \mathbf{Z}^{T}\mathbf{A}\,\mathbf{Z}.
\end{eqnarray}
Here $\mathbf{I}$ is the identity matrix of size ${{\cal N}_p\times
  {\cal N}_p}$, and $\mathbf{Z}$ is a tall and skinny matrix  of size
${{\cal N}_p\times r}$ and rank $r \, (\ll {\cal  N}_p)$.  Matrix $\mathbf{Z}$ is referred to as a deflation
subspace matrix, while matrix $\mathbf{E}$ is called  a coarse operator.
As $\mathbf{A}$ is SPD, so is
$\mathbf{E}$.  The size of $\mathbf{E}$ is $r \times r$, 
and its direct factorisation can be easily computed.
The columns of $\mathbf{Z}$ are linearly independent and are selected in such a way
that they span the subspace, referred to as the
deflation subspace, which is to be projected out from any vector it is applied to. That this is
indeed the case 
can be seen noting that $\mathbf{R} \mathbf{A} \mathbf{Z} = \mathbf{0}$.
When applied to the case at hand, the deflation
subspace is defined to contain
the eigenvectors corresponding to small eigenvalues of $\mathbf{A}$.
As a result, all these eigenvalues are replaced by zeros in the
spectrum of $\mathbf{R}\,\mathbf{A}$, while all the others remain unchanged.
In the exact precision
arithmetic, $\mathbf{R}$ would potentially be a very
efficient preconditioner, as all the steps of any iterative CG-like
solver would be orthogonal to the null space of the preconditioned
matrix, $\mathbf{R}\,\mathbf{A}$.  However, in the finite precision
arithmetic, this may not be the case, and the zero eigenvalues are
often as bothersome as the small ones due to numerical errors. 
Another practical problem is that the dimension of the deflation subspace, given by the number of columns of $\mathbf{Z}$, 
is typically limited for computational reasons. Hence, preconditioners based on deflation are most
effective when the system matrix has only few small and outlying eigenvalues. 
 
Both these issues can be overcome by combining a deflation preconditioner with another one.
Such  combined constructions are referred to as two-level preconditioners. 
There are numerous prescriptions, both additive or multiplicative, in the literature proposing how to combine
two preconditioners. In our work, we use
the proposal of~\citet{TANG:2009} and combine them together with the standard and deflation preconditioners as follows:

\begin{eqnarray} \label{eq:2}
\mathbf{M}_{2lvl} & := & \mathbf{M}_{BD}\,\mathbf{R}\,  + \, \mathbf{Z}\,\mathbf{E}^{-1}\mathbf{Z}^{T}\\
 & = & \mathbf{M}_{BD}\left(\mathbf{I} -\mathbf{A}\,\mathbf{Z}\,\mathbf{E}^{-1}\,\mathbf{Z}^{T}\right) + \mathbf{Z}\,\mathbf{E}^{-1}\,\mathbf{Z}^{T}.\, \nonumber
\end{eqnarray}
This corresponds to the "Adapted Deflation Variant 1" method (A-DEF1)
in \citet{TANG:2009}. We note that this may not be the most obvious choice from a purely theoretical perspective, see Appendix~\ref{sec:alt_prec}.
However, this is the one, which has proven to be the most robust and efficient in our numerical experiments, and for this reason, we have adapted it in this
work.

We note that this new, two-level preconditioner indeed resolves the two problems mentioned before. First,
\begin{eqnarray}
\mathbf{M}_{2lvl}\,\mathbf{A}\mathbf{Z} \, = \, \mathbf{Z},
\end{eqnarray}
and therefore the two-level preconditioner rescales all the eigenvalues of $\mathbf{A}$ contained in the deflation subspace defined by $\mathbf{Z}$ to unity. Second,
for $\mathbf{Y}^t\mathbf{A}\,\mathbf{X}\,=\,\mathbf{0}$,
\begin{eqnarray}
\mathbf{M}_{2lvl}\,\mathbf{A}\mathbf{Y} \, = \, \mathbf{M}_{BD} \, \mathbf{A} \, \mathbf{Y}.
\end{eqnarray}
That is, the action of the two level preconditioner, $\mathbf{M}_{2lvl}$, on the eigenvectors of $\mathbf{A}$ orthogonal to the deflation subspace in the sense of the A-norm is
equivalent to that of the standard preconditioner, which we have seen in Figure~\ref{fig:eigen_values}
effectively shifts the large eigenvalues towards one. If $\mathbf{Z}$ is defined to include all
 small eigenvalues of $\mathbf{A}$, the two-level preconditioner will therefore shift both the small and large eigenvalues of  $\mathbf{A}$  to the vicinity of one. In practice, the size of the deflation
 subspace is limited, so this can be achieved only partially. 
The challenge, therefore, is in a construction of $\mathbf{Z}$ that ensures that the smallest eigenvalues, which are potentially the most troublesome eigenvalues from the point of view of the iterative method convergence, are included and which should be numerically efficient so the performance of this approach is competitive with the standard method.
 We discuss two different proposals for the latter task in the next Section.

%
%

\subsection{Deflation subspace operator}\label{sec:coarse_space} 

 In the ideal case, the columns of the deflation subspace matrix, $\mathbf{Z}$,
should be made of the eigenvectors corresponding to the smallest
eigenvalues of the preconditioned system matrix,
$\mathbf{M}_{BD}\,\mathbf{A}$.  However these are typically too
expensive to compute and, instead, one needs to resort to some
approximations.

There are  two broad classes of suitable approximation schemes, which
are either {\em a priori}, and therefore using only the initial knowledge
of the problem that we want to solve to construct $\mathbf{Z}$, or {\em a
  posteriori}, which resorts to some explicit precomputation performed ahead of the
actual   solution.
 The precomputation can rely on prior
solutions of analogous linear systems solved via some iterative
algorithm and some simpler preconditioner, as, for instance, $\mathbf{M}_{BD}$
in our case.  The preconditioners of this type are clearly
useful if the map-making problem with the same system matrix needs to be solved multiple
times for different right hand sides, so the precomputation cost quickly becomes irrelevant.  This is
indeed the case in many practical situations, be it extensive Monte
Carlo simulations or posterior sampling algorithms, such as the Monte Carlo
Markov Chain approaches, which both are frequently used in CMB
data analysis.

In~\citet{midas_sc12}, we proposed an {\em a priori} two-level
preconditioner suitable for total intensity observations of the
CMB.  Below, we first generalise this preconditioner to
polarisation-sensitive observations and then introduce a new {\em a posteriori} preconditioner.

\subsubsection{A priori construction}\label{sub:a_priori_construction} 

For the total intensity case, our {\em a priori} two level preconditioner is
based on the deflation subspace built to include a pixel domain vector of ones,
$\mathbf{1_p}$. This is because the vector of ones is in the near
nullspace of the system matrix, whenever long-term time
domain correlations are present.  To demonstrate this, let us
consider the typical power spectrum of such noise, as given in
equation~\eqref{eq:fknee}. This spectrum displays a
characteristic "$1/f$" behaviour at the low-frequency end, which
results in a significant excess of power in this regime.
Though more realistic noise spectra typically flatten out at very
low frequencies, instead of continuing to increase as $1/f$, the power excess is nevertheless 
present, as the flattening happens at frequencies much lower than $f_{knee}$.
As the noise spectrum corresponds to
the eigenvalues of the noise correlation matrix, $\mathbf{N}$, as in 
equation~\eqref{eq:N_def}, it is apparent that the eigenvalue of the
zero frequency mode is significantly larger than the majority of the
other eigenvalues corresponding to the high frequency plateau of the
spectrum.  Consequently, the zero frequency mode given by a vector of
ones in the time-domain, $\mathbf{1_t}$, is in the near nullspace of
the inverse time-domain correlation matrix, $\mathbf{N}^{-1}$, that is
$\mathbf{N}^{-1}\,\mathbf{1_t} \, \simeq \, 0$.  Hence, given that
\begin{eqnarray}
\mathbf{A}\,\mathbf{1_p} = 
\mathbf{P}^t\,\mathbf{N}^{-1}\,\mathbf{P}\,\mathbf{1_p}  = 
\mathbf{P}^t\,\mathbf{N}^{-1}\,\mathbf{1_t} \simeq 0,
\label{eq:AsingVect}
\end{eqnarray}
 the pixel domain vector of ones, $\mathbf{1_p}$, is expected to be 
in the near nullspace of
$\mathbf{A}$.  We can, thus, expect that including this vector in the
deflation subspace should result in significant gains in the solver's
convergence, as is indeed borne out by our results discussed in section
\ref{sec:numerical_experiments}.
 
We can strive to further accelerate the convergence by
employing a richer deflation subspace matrix, $\mathbf{Z}$.  
In the approach proposed here, we capitalise on the observation that the instrumental noise, $\vec{\breve{n}},$ is
piece-wise stationary. Let us assume that we have a disjoint $K+1$ stationary intervals,
$\vec{d}^{0}, \ldots, \vec{d}^{K}$.  Each interval, $\vec{d}^j$, is
associated by construction with a unique time domain noise correlation
matrix, $\mathbf{T}^j$.
The deflation matrix, which we denote hereafter as $\mathbf{Z}_S$, is built by assigning each of its
columns to one of the piece-wise stationary time
intervals~\citep[][]{midas_sc12}, such that the $j$th column
corresponds to the $j$th stationary interval, $\vec{d}^j$. In the case
of the total intensity observations for a given pixel, $p$, the
elements in the $j$th column define the fraction of the observations
of this pixel performed within the $j$th period, $s_p^j$, as compared
to all its observations, $s_p$,

\begin{eqnarray}
\label{eq:strofZ}
\begin{array}{rclcl}
\mathbf{Z}_S^{tint} & := &
\begin{pmatrix}\smallskip
	{\displaystyle\frac{s^{0}_{0}}{s_{0}}} & {\displaystyle\frac{s^{1}_{0}}{s_{0}}} & {\displaystyle\cdots} & {\displaystyle\frac{s^{K}_{0}}{s_{0}}} \\
         {\displaystyle \frac{s^{0}_{1}}{s_{1}}} & {\displaystyle\frac{s^{1}_{1}}{s_{1}}} & {\displaystyle\cdots} & {\displaystyle\frac{s^{K}_{1}}{s_{1}}} \\
	{\displaystyle\vdots} 			         & {\displaystyle\vdots} 			   & 	                     	      & {\displaystyle\vdots}			         \\
	{\displaystyle\frac{s^{0}_{p}}{s_{p}}} & {\displaystyle\frac{s^{1}_{p}}{s_{p}}} & {\displaystyle\cdots} & {\displaystyle\frac{s^{K}_{p}}{s_{p}}}
\end{pmatrix}. 
\end{array}
\end{eqnarray}  
We note that each row of $\mathbf{Z}_S^{tint}$ represents some partition of unity as $\sum_j\,\mathbf{Z}^{tint}_S(p,j) \, = \, 1$.

In experiments with polarisation, the extra entries of each column, corresponding to Q and U signals, are simply set to $0$, so the deflation subspace matrix for the polarised cases is given by:
\begin{eqnarray}
\label{eq:strofZ_polar}
\begin{array}{cccc}
	\left(\mathbf{Z}_S^{pol}\right) & := &
	\begin{pmatrix}\smallskip
		{\displaystyle\frac{s^{0}_{0}}{s_{0}}}    & {\displaystyle\frac{s^{1}_{0}}{s_{0}}} & {\displaystyle \cdots} & {\displaystyle\frac{s^{K}_{0}}{s_{0}}} \\
		 {\displaystyle 0} & {\displaystyle \cdots} & {\displaystyle \cdots} & {\displaystyle 0}\\
		 {\displaystyle 0} & {\displaystyle \cdots} & {\displaystyle \cdots} & {\displaystyle 0}\\
		 {\displaystyle \frac{s^{0}_{1}}{s_{1}}}  & {\displaystyle\frac{s^{1}_{1}}{s_{1}}} & {\displaystyle \cdots} & {\displaystyle\frac{s^{K}_{1}}{s_{1}}} \\
		 {\displaystyle 0} & {\displaystyle \cdots} & {\displaystyle \cdots} & {\displaystyle 0}\\
		 {\displaystyle 0} & {\displaystyle \cdots} & {\displaystyle \cdots} &{\displaystyle 0}\\
		  {\displaystyle \vdots} & {\displaystyle \vdots} & {\displaystyle \vdots} &{\displaystyle \vdots}\\
		  {\displaystyle\frac{s^{0}_{p}}{s_{p}}} & {\displaystyle\frac{s^{1}_{p}}{s_{p}}}  & {\displaystyle \cdots} & {\displaystyle\frac{s^{K}_{p}}{s_{p}}} \\
		  {\displaystyle 0} & {\displaystyle \cdots} & {\displaystyle \cdots} &{\displaystyle 0}\\
		  {\displaystyle 0} & {\displaystyle \cdots} & {\displaystyle \cdots} & {\displaystyle 0}
	\end{pmatrix}. 
\end{array}.
\end{eqnarray}  

This choice essentially implies that we apply no correction to the
eigenvectors of the system matrix, which have a non-zero component
in the part of the space corresponding to the polarised degrees of
freedom. 
This seems justified given that these are degrees of freedom related to the temperature, which
are expected to lead to the smallest eigenvalues. We note for instance that the deflation subspace
defined in this way does not include the unit vector, that is a vector of ones for all pixels and all three Stokes parameters
but it includes only a vector of ones for the temperature pixels and zero otherwise.
This is consistent with equation~\eqref{eq:AsingVect}, which implies that this is the latter not the former vector, which generally is in the near null 
space of $\mathbf{A}$ whenever $1/f$ noise is present. For the unit vector, $\mathbf{1}_{3p}$, and for polarisation-sensitive observations, typically
$\mathbf{P}\,\mathbf{1}_{3p}\,\ne\,\mathbf{1_{t}}$, while equality is assumed in the derivation of equation~\eqref{eq:AsingVect}.
We note that this choice is efficient in improving the condition number of the system matrix, as can be seen
in Figure~\ref{fig:eigen_values}, and we test its impact on the convergence of the iterative solvers in Section~\ref{sec:results_and_discussion}.

The deflation subspace matrix, $\mathbf{Z}_S$, constructed in this section can be
rather sparse, reflecting that only a small subset of all pixels is typically observed within each stationary period. The sparsity
is therefore uniquely given by the definition of the stationary
periods and the structure of the pointing matrix and
can be predicted ahead of time and used to implicitly construct the preconditioner, which is then applied at each
step of the iterative solver.  As the number of columns, $r$, of
the operator $\mathbf{Z}_S$ is tied to the number of stationary
periods, it may seem that the size of $\mathbf{A}$ is uniquely
predetermined by the data set properties. However, though not
completely arbitrary, the number of columns of $\mathbf{A}$ can always be reduced by combining some of the stationary periods
together. As we discuss in Section~\ref{sub:coarse_space_impact},
at least in some cases, such an approach can ensure satisfactory
performance at lower numerical cost with fewer columns of
$\mathbf{Z}_S$. We emphasise that the reduction of the size of $\mathbf{Z}_S$
must be performed consistently, and in particular the sum of all
elements for each nonzero row of $\mathbf{Z}_S$ has to remain equal to
$1$ after such an operation is performed. Hereafter, unless stated
explicitly, we always take $r=K$.

\subsubsection{A posteriori construction}\label{sub:a_posteriori_construction}  

In this case, the deflation subspace matrix, $\mathbf{Z}$, is constructed based
on some suitably devised precomputation, which typically yields  approximate eigenvectors
of the preconditioned system matrix, $\mathbf{M}_{BD}\,\mathbf{A}$.  A particularly interesting
option, which we focus on here, is when the precomputation consists in solving a similar linear
system to the actual one and featuring the same system matrix, $\mathbf{A}$, but potentially 
different right-hand side, $\vec{b}'$,
\begin{eqnarray}
  \label{eq:mat_sequence}
  \mathbf{A}  \vec{x}' = \vec{b}'.
\end{eqnarray}
Any iterative solver of such systems based on the
Krylov subspace methods (see Appendix~\ref{app:aprox_eigenvector})
 internally derives information about the spectral
properties of the system matrix.  This information can, therefore, at least in principle, be stored and then re-used in building
the deflation subspace matrix (denoted hereafter $\mathbf{Z}_D$) of a two-level preconditioner,
which could be subsequently applied in the following solves of the
same system.  We could expect that this approach leads to a two-level preconditioner, which generally is more efficient than any
{\em a priori} construction, as the coarse space operator
constructions is performed using estimates of the true eigenvectors of
the matrix.

There are many specific proposals of these constructions in the
numerical linear algebra literature, 
which are specialised for either GMRES \citep{Morgan_GMRES,
  kharchenko95:_eigen_trans_based_precon_for, Chapman96deflatedand,
  Saad:2000:DVC, Parks:2006:RKS} or the conjugate gradient method
\citep{ACG_SIAM,Risler:2000:IAA,Gosselet:2003:SRK}.  For more details,
we refer the reader to Appendix~\ref{app:aprox_eigenvector} and the
original papers, while we just list the main common steps of
these algorithms here. These steps are as follows:
\begin{enumerate}
 \item Solve the initial linear system (or systems)
   $\mathbf{M}_{BD}\,\mathbf{A}\,\vec{x}'\,=\,\mathbf{M}_{BD}\,\vec{b}'$
   using a Krylov subspace iterative method. As a
   by-product, we obtain a spectral decomposition of the respective
   Krylov subspace, see equation~\eqref{eq:krylov_block_prec}.
 \item Derive approximations of the eigenvalues and eigenvectors of
   $\mathbf{A}$ by employing the Ritz eigenvalues approach,
   see equation~\eqref{eq:approxEigen}.
 \item Select the Ritz eigenvectors, which correspond to
   eigenvalues with real parts smaller than some given threshold,
   $\epsilon_{tol}$. (Note that we describe here a more general setting, which
   does not assume that the matrices are SPD).
 \item Construct the deflation subspace matrix as
   $\mathbf{Z}_D:=|\vec{V}_1|\vec{V}_2|\ldots|\vec{V}_{r}|$, where
   $(\vec{V}_1,\ldots,\vec{V}_{r})$ denote $r$ eigenvectors selected
   in the previous step.
\item Construct our two-level preconditioner using $\mathbf{Z}_D$.
\end{enumerate}

Unlike the matrix $\mathbf{Z}_S$ discussed earlier,
$\mathbf{Z}_D$ will typically be dense.  However, the number of its
columns $r$ could essentially be set arbitrarily and be typically determined by a suitable choice of the tolerance threshold,
$\epsilon_{tol}$.
 
We also note that in principle one could imagine that the deflation subspace operator
constructed for the current run is subsequently updated with the new
information obtained from it before being used for the next run if many map-making runs have to be performed.
Thus, it can be potentially used to progressively construct a more
efficient coarse operator. From our experience, the additional
benefits of such an approach are very quickly becoming minor and come
at the cost of a significant extra complexity of the implementation.
We have, therefore, not considered this option in this work.
 
\subsection{Cost of application and storing $\mathbf{M}_{2lvl}$}
\label{sub:cost2level} 


The two-level preconditioners, as defined in equation~\eqref{eq:2},
are in general dense.  Therefore, in the algorithms proposed here, they are not and not can be constructed explicitly. Instead, we precompute only the matrix
 $\mathbf{A}\mathbf{Z}$ and the coarse operator, $\mathbf{E}$, which we store in the computer
 memory. We then apply $\mathbf{M}_{2lvl}$  to a vector as in equation~\eqref{eq:2}, performing the matrix vector operations from left to right.

The details of our implementation are given in Appendix~\ref{sec:parall-constr-coarse}. Once $\mathbf{A}\,\mathbf{Z}$ and $\mathbf{E}$ are available they demonstrate that the operations required by the two-level preconditioner are performed within either deflation or pixel space and are, therefore, subdominant as 
compared to the time-domain operations,  such as noise-weighting, which is involved in the
products of the system matrix by a vector and, therefore, needs to be repeated on every iteration.
We can, therefore expect that the extra computational cost introduced by these preconditioners are manageable.  We confirm this expectation by means of numerical experiments in Section~\ref{sec:results_and_discussion} and Appendix~\ref{sec:parall-constr-coarse}.


\section{Numerical experiments}\label{sec:numerical_experiments} 

\begin{figure*}[!ht]
	\subfloat[]{
	\begin{minipage}[c][1.25\width]{0.22\textwidth}
		\centering%
\label{fig:simple_hits_rectangular}\includegraphics[width=\textwidth]{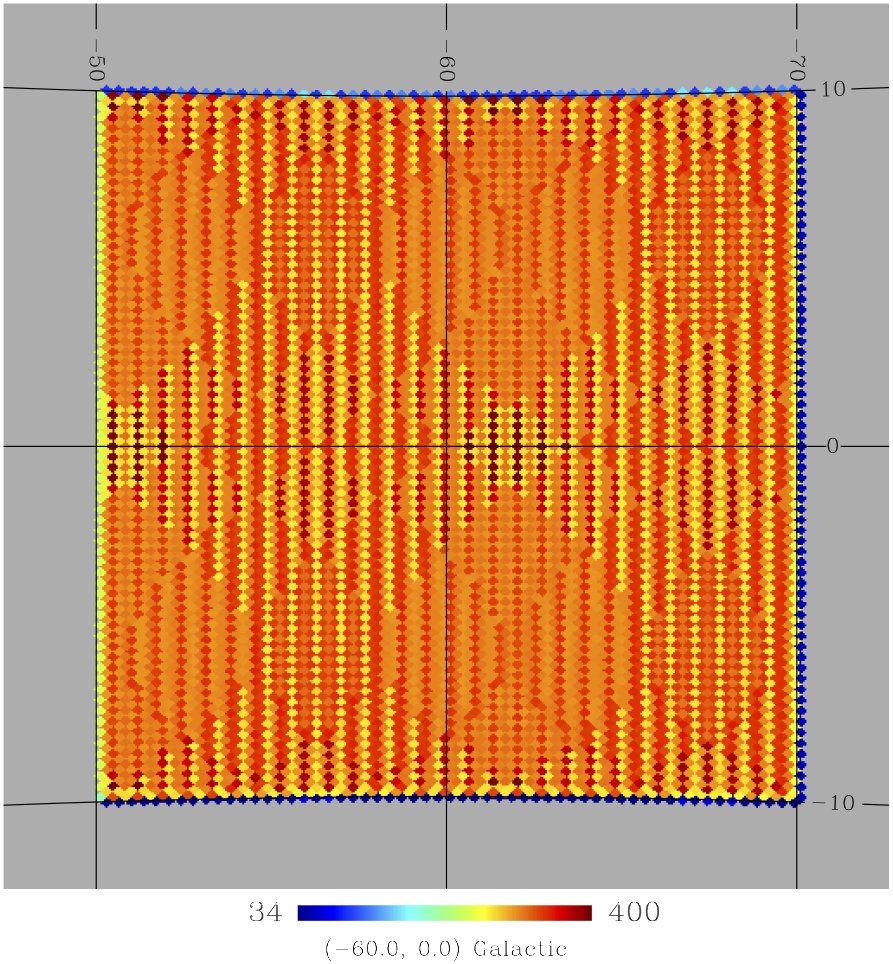}
\end{minipage}}
\hfill
\subfloat[]{
	\begin{minipage}[c][1.25\width]{0.22\textwidth}
	   \centering%
	   \label{fig:simple_hits_circles}\includegraphics[width=\textwidth]{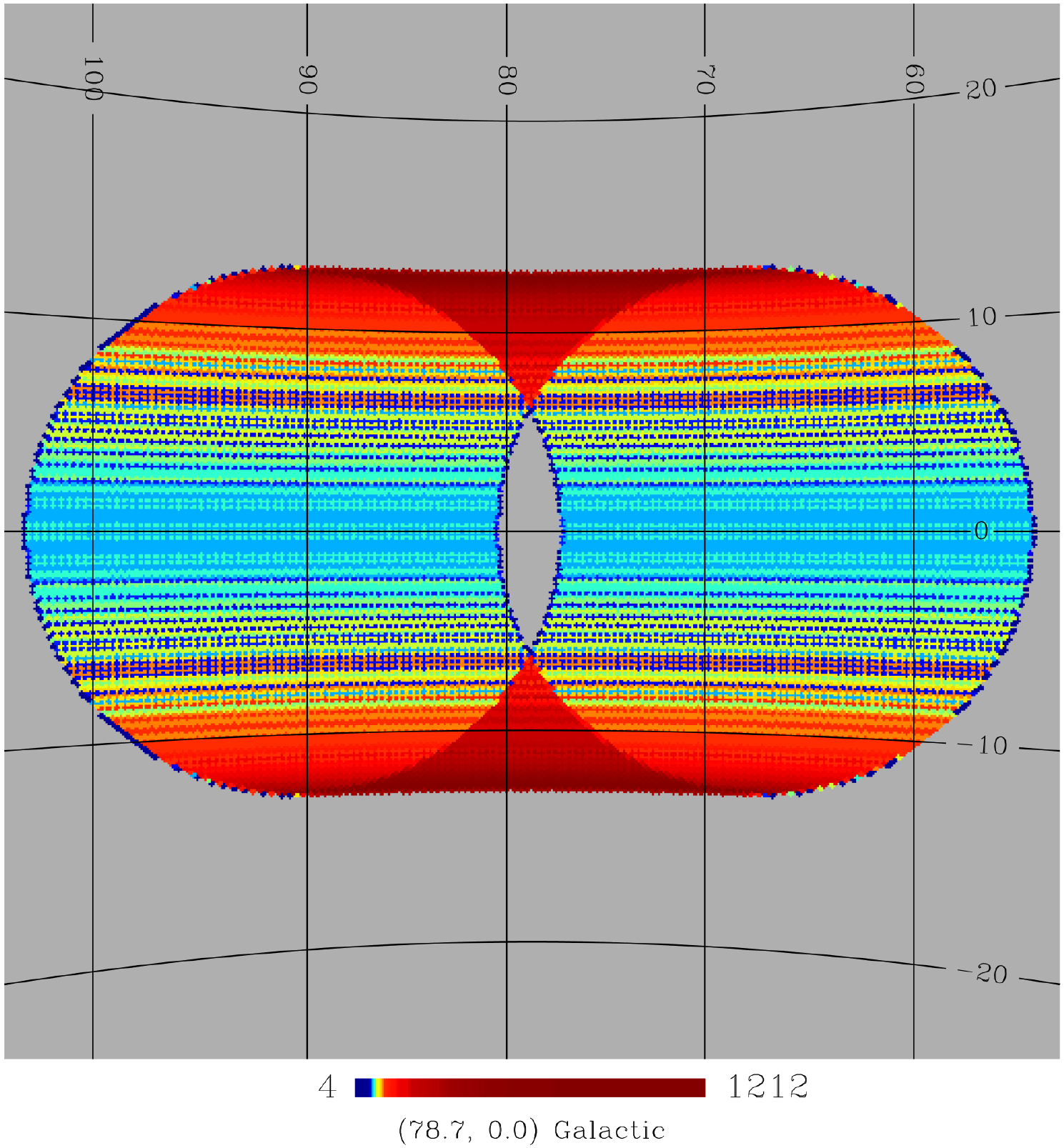}
	\end{minipage}}
	\hfill
	\subfloat[]{
\begin{minipage}[c][0.6\width]{0.45\textwidth}
   \centering%
   \label{fig:tk_hits}\includegraphics[width=\textwidth]{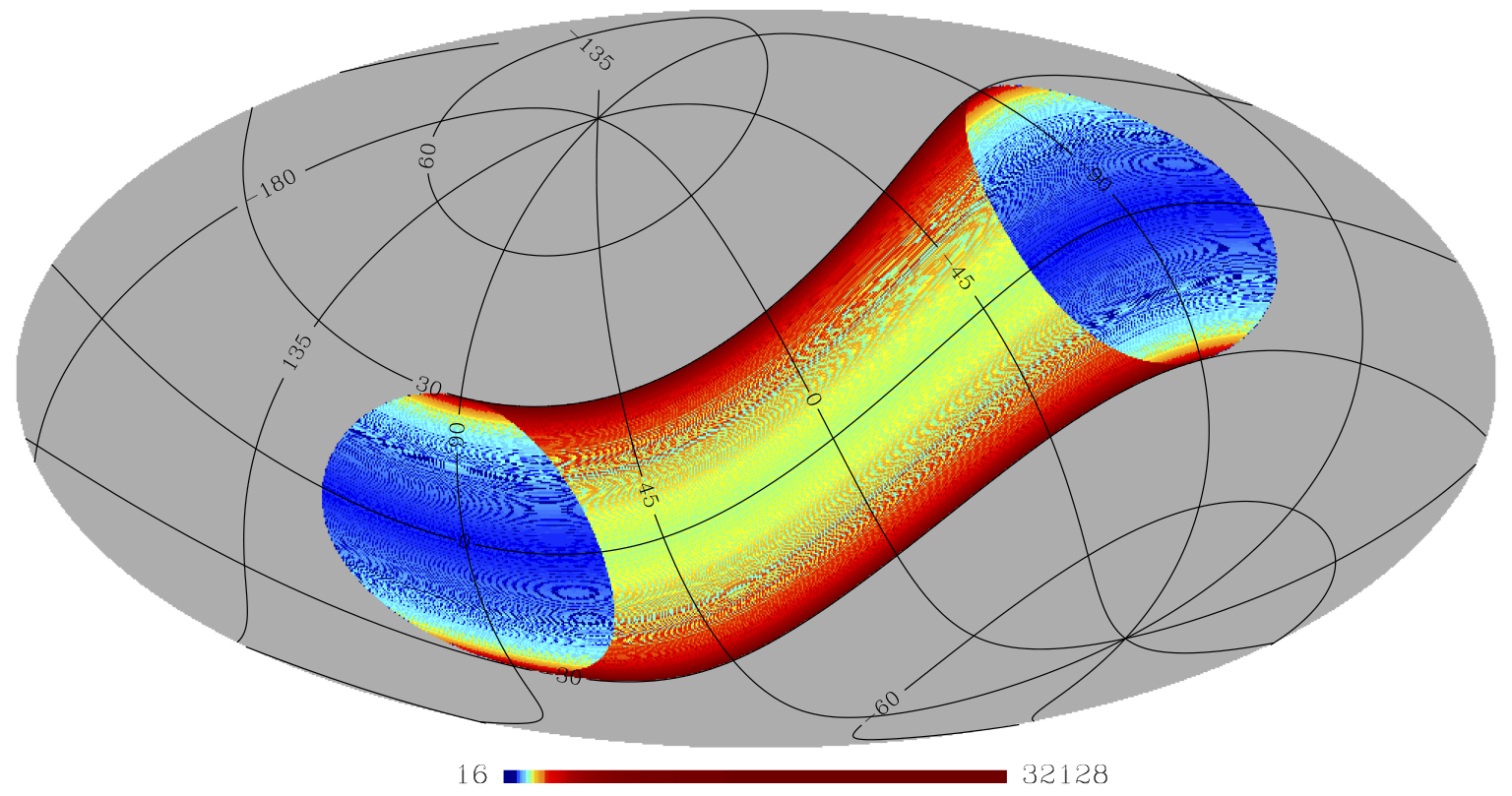}
\end{minipage}}
\\ 
\subfloat[]{
	\begin{minipage}[c][0.6\width]{0.45\textwidth}
	   \centering%
	   \label{fig:toeplitz}\includegraphics[width=\textwidth]{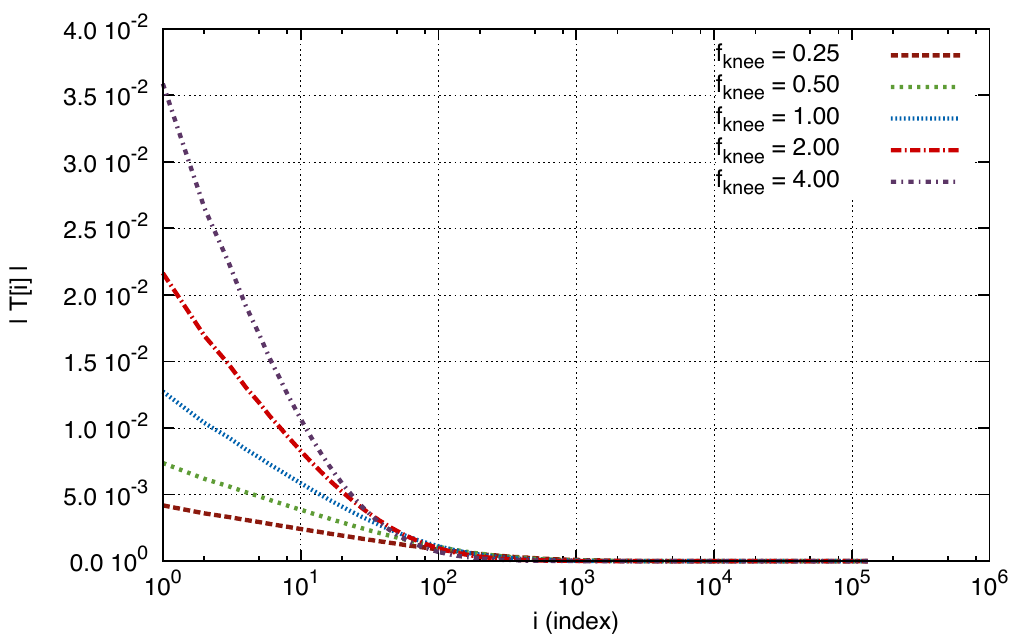}
	\end{minipage}}
	\hfill
	\subfloat[]{
		\begin{minipage}[c][0.6\width]{0.45\textwidth}
		   \centering%
		   \label{fig:ref_sky}\includegraphics[width=\textwidth]{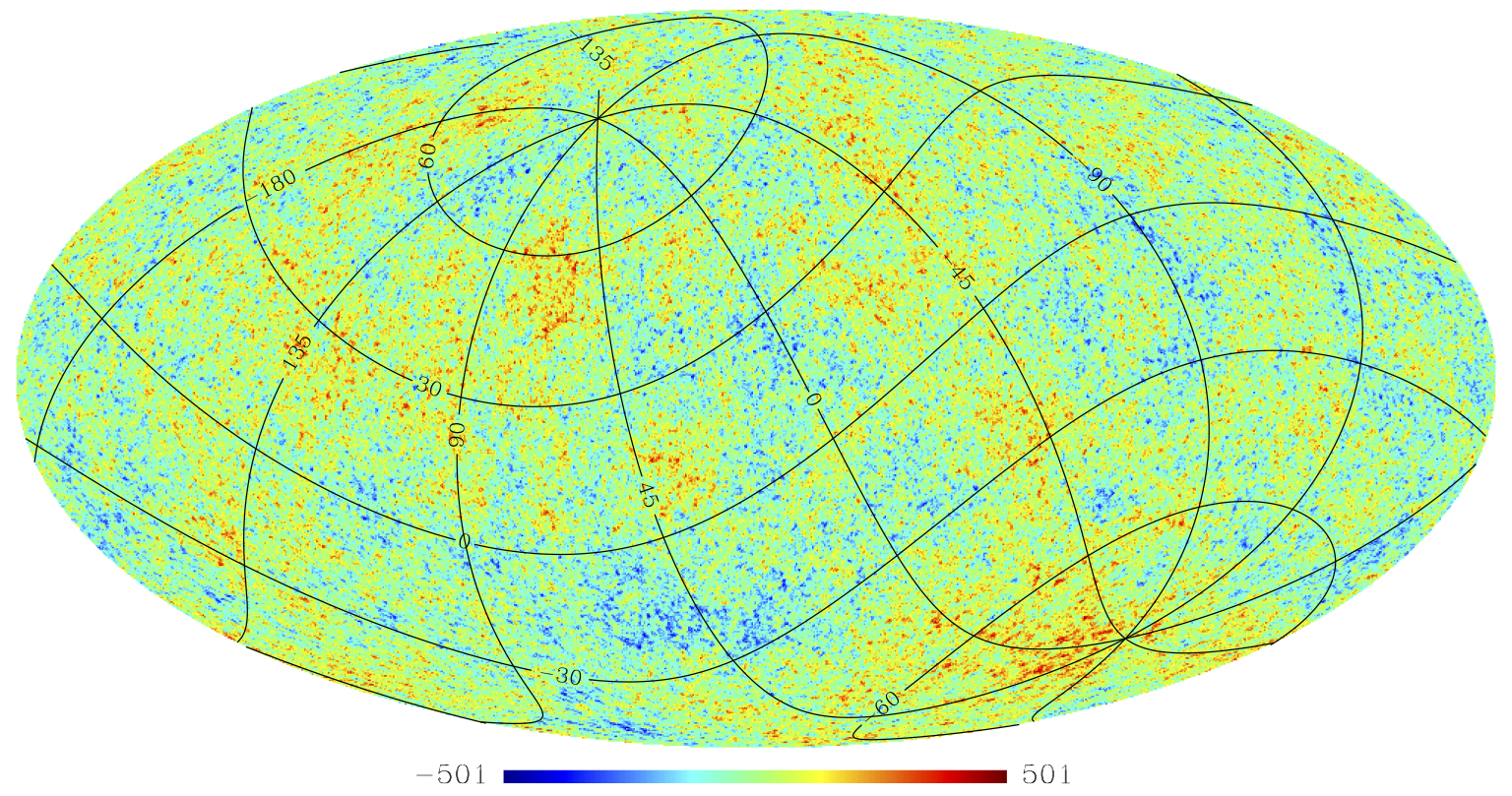}
		\end{minipage}}

\caption{Visualisation of the input information used for the simulations (see Sect.~\ref{sec:simulations} for more details). {\bf Upper row:} panels (a-c), hit maps obtained for the grid-like, small circle, and big circle scans, respectively. {\bf Bottom row:} (d) noise correlations assumed for the stationary intervals and corresponding to different values of the knee frequency $f_{knee}$, (the diagonal, highly dominant element is not shown to emphasise the off-diagonal correlations), (e) an example of the total intensity map of the CMB fluctuations.}
\end{figure*}

In this section, we present a number of numerical experiments
demonstrating that PCG preconditioned by the two-level preconditioners
provide an efficient and robust iterative solver for the map-making
problem. The numerical experiments have been designed with the
following main objectives in mind:
\begin{enumerate}
\item To demonstrate that the number of
  iterations needed to converge that is to reach a given pre-defined
  magnitude of the residuals, is smaller in the case of the
  two-level preconditioner than in the case of the standard one. 
\item To demonstrate that the two-level
  preconditioner also leads to important savings in terms of time to
  solution.
\item To demonstrate that the two-level
preconditioner can be implemented for massively parallel computers and 
in a scalable manner. That is, that it is capable of maintaining its performance and, in
particular, fulfils the two first objectives, with an increasing 
number of processors. This is necessary given the volumes of current and future CMB data sets.
\end{enumerate}
In the following, we discuss the efficiency of the two-level
preconditioners with respect to these objective for a range of
simulated CMB-like observations with realistic characteristics. In
particular, we consider different scanning patterns, sizes of the
data set in time- and pixel- domains, noise correlation lengths,
etc. We also discuss the role of the bandwidth
assumption, which is typically imposed on the Toeplitz matrices used to describe
properties of stationary time domain noise, as seen in equation~\eqref{eq:N_def}.  As
the reference for all the runs described below, we use the PCG solver
with the standard block-diagonal preconditioner,
$\mathbf{M}_{BD}$.

\subsection{Simulations}\label{sec:simulations} 

The numerical experiments use simulated CMB data sets with characteristics inspired by actual CMB experiments but are kept simple to permit inferences about the impact of their different parameters on the performance of the method. In general, the time ordered data (TOD) is simulated by generating a list of telescope pointings produced for a selected scanning strategy and assigning the respective signals to it, $\mathbf{i}_p, \mathbf{q}_p$, and $\mathbf{u}_p$, as read off from the simulated CMB maps. The recovered signals are combined together as in equation~\eqref{eq:signal_with_polarization}, added to simulated instrumental noise $\vec{\breve{n}_{t}}$, and stored as a time domain vector, $\vec{d}_t$.

\subsubsection{Scanning strategies}\label{sub:scanning_strategies} 

We employ three simplified scanning strategies in our simulations, as listed in Table~\ref{tab:cases}. They correspond to either small-scale or nearly full-sky observations, have different sizes in the time and pixel domains, and have various levels of redundancies present in the scan, as quantifiable by a different distribution of the number of revisits of  sky pixels, or  a different distribution of the attack-angles in pixels. The studied cases are
\begin{enumerate}
	\item A grid-like scan
	in which a patch of $20^{\circ} \times 20^{\circ}$ deg on the sky is
	observed in turn either horizontally or vertically, as seen in Figure \ref{fig:simple_hits_rectangular}.
	\item A scan made of $128$ circles on the sky with the diameter of
	$15^{\circ}$ deg and centres located along one of the celestial sphere's great circles, as seen in Figure \ref{fig:simple_hits_circles}. 
	Each circle is scanned four times
	before the experiment moves to the another one. 
	\item A scan pattern made of $2048$ densely
	crossing circles, as shown in Figure~\ref{fig:tk_hits}, which are
	scanned one-by-one, consecutively, from left to right.  The circles
	have the opening angle of $30^\circ$. The time spent on each circle is
	assumed to be the same. Each circle is scanned $16$ times and sampled
	$\approx10^6$ times before the experiment switches to another one. As
	a result of such a procedure, we have a vector of more than two
	billion observation directions. This scan reflects some of the basic properties of a satellite-like observation.
\end{enumerate}

\begin{table}
\begin{center}
\caption{Size of the problem for the different cases analysed.}
\label{tab:cases}
\begin{tabular}{clccc}
\hline
&  & \# of stationary  & \# of & \# of  \\
Case & Pattern  & intervals & pixels & time samples\\
\hline
1 & Grid scan & 1 & 7762  &$\simeq 1.0 \times 10^6$ \label{case:1} \\
2 & Small circles & 1 & 20585 & $\simeq 2.0 \times 10^6$ \label{case:2} \\
3 & Big circles & 2048 & 998836 & $\simeq 2.0 \times 10^9$ \label{case:3} \\
\hline
\end{tabular}                                                                                           
\end{center}
\end{table}

\subsubsection{Polariser}\label{sub:polariser} 
We specify a dependence of the orientation of the polariser, $\phi_{t}$, as defined in equation~\eqref{eq:signal_with_polarization}, on time, assuming three scenarios.
\begin{enumerate}
	\item {\bf Fast polariser rotation:} The polariser is rotated by $45^{\circ}$ deg from one to another pointing along the scan direction.
	\item {\bf Medium rotation:} We change the polariser angle by $45^{\circ}$ deg after each scan of one circle (for the circular scans) or after changing of direction of scanning (for the grid-like scan).
	\item {\bf Slow rotation:} In this case we repeat the full scan four times before changing the polariser angle by $45^{\circ}$ deg. As a consequence, the time domain is four times larger than the ones listed in Table~\ref{tab:cases}.
\end{enumerate}

\subsubsection{CMB signal}\label{sub:cmb_signal} 
Theoretical power spectra have been calculated using the CAMB package\footnote{\url{http://camb.info}}~\citep{CAMB:2011} assuming the concordance model parameters~\citep{PlanckPars2013}. They have subsequently been convolved with a fiducial symmetric Gaussian instrumental beam of FWHM $10$ arcminutes and used to  simulate a Gaussian realisation of the CMB map using the \emph{synfast} routine from a HEALPix~\citep{HEALPix} with a HEALPix resolution parameter set to ${\cal N}_{side} = 512$. 

\subsubsection{Noise}\label{sub:noise} 
We have simulated the noise contribution, $\vec{\breve{n}_{t}}$, as a Gaussian realisation of a piece-wise stationary noise composed of an uncorrelated white noise and correlated $1/f$ components with the  power spectrum density for each of the stationary pieces that are that are parametrized in the usual manner as
\begin{eqnarray}\label{eq:fknee}
	P(f) = \sigma^{2}\,t_{samp}\left( 1 + \left( \frac{f_{knee}}{f} \right)^2 \right).
\end{eqnarray}
Here, $\sigma^2$ defines the white noise level,  $t_{samp}$ is the sampling interval, and $f_{knee}$ -- the knee frequency of the spectrum. 
For definiteness, we used $\sigma^2 = 8.8\,10^{-10}$K$_{CMB}^2$ for a single observation noise variance -- a value roughly corresponding to a typical noise level of current bolometric detectors.

For the two smaller scans, cases 1 and 2 in Table~\ref{tab:cases}, we have assumed that the noise is stationary over the entire period of the observation,
while stationary intervals correspond to the time spent scanning each circle for the largest experiment (case 3). This implies that 
noise correlations are then present only between the samples corresponding to the same circle, and
the resulting (inverse) noise covariance, as seen in equation~\eqref{eq:N_def}, is composed of as many diagonal blocks, as circles, that is up to $2048$.          
This is in contrast  with cases 1 and 2, where there is only one Toeplitz block.
 
For any stationary interval, noise correlations may exist between any two of its samples however distant they are. 
This would correspond to full
dense Toeplitz blocks of the (inverse) noise correlation matrix,
$\mathbf{N}^{-1}$.  It is, however, frequently assumed that the
correlations are zero if the time lag between the samples exceeds some correlation 
length, $\lambda_T$. If this is indeed the case, the Toeplitz blocks of $\mathbf{N}^{-1}$ 
are band-diagonal with a half bandwidth
set by $\lambda_T$. In our experiments for definiteness, we set this parameter to be
$\lambda_{T} = 2^{13}$, unless it is explicitly stated otherwise. We discuss the role and importance of this parameter on the
efficiency of the iterative solvers in
Section~\ref{sec:results_and_discussion}.

\subsection{Experimental setup}\label{sub:experimental_setup} 

We evaluate performance of our two-level solver on NERSC's supercomputer Hopper\footnote{ \url{http://www.nersc.gov/nusers/systems/hopper2/}}, based on a Cray XE6 system.  Hopper is made up of  $6384$ compute nodes with each node composed of two twelve-cores AMD MagnyCours ($2.1$ GHz), with a peak
performance of $201.6$ Gflops per node and $1.288$ Peta-flops for the
entire machine. The nodes are connected with Cray's Gemini high performance interconnect
network.

For the tests we use our own parallel code written in C++, which uses
a multithreaded version of Intel's Math Kernel Library for calculation of the Fast Fourier Transforms and for dense linear algebra. We use
Intel Compiler version 12.1.4.

The code is fully parallelised using a distributed-memory programming model with help of the MPI (Message Passing Interface)
library.
In addition, some routines in our code and the {\tt MKL} library used in our implementation allow us to exploit multithreading. This is
done by assigning each MPI (distributed memory) process to a single multicore processor and then by capitalising on its multicore
structure with OpenMP threads, which is managed via calls to the OpenMP library. On Hopper, we use four MPI processes 
per compute node and six OpenMP threads per MPI process. Consequently, a run using $2048$ MPI processes typically uses
up to $2048 \times 6 = 12,288$ cores. Details of our implementation and the assumed data layout are given
in section~\ref{sec:parall-constr-coarse}.

 

\section{Results and discussion}\label{sec:results_and_discussion} 

\subsection{Convergence comparison}

\begin{figure*}[tbp]
	\centering
	\subfloat[Polarisation case 1]{%
		\begin{minipage}[c][0.39\width]{0.84\textwidth}
		\centering
		\label{fig:convergence_pc1}\includegraphics[width=\textwidth]{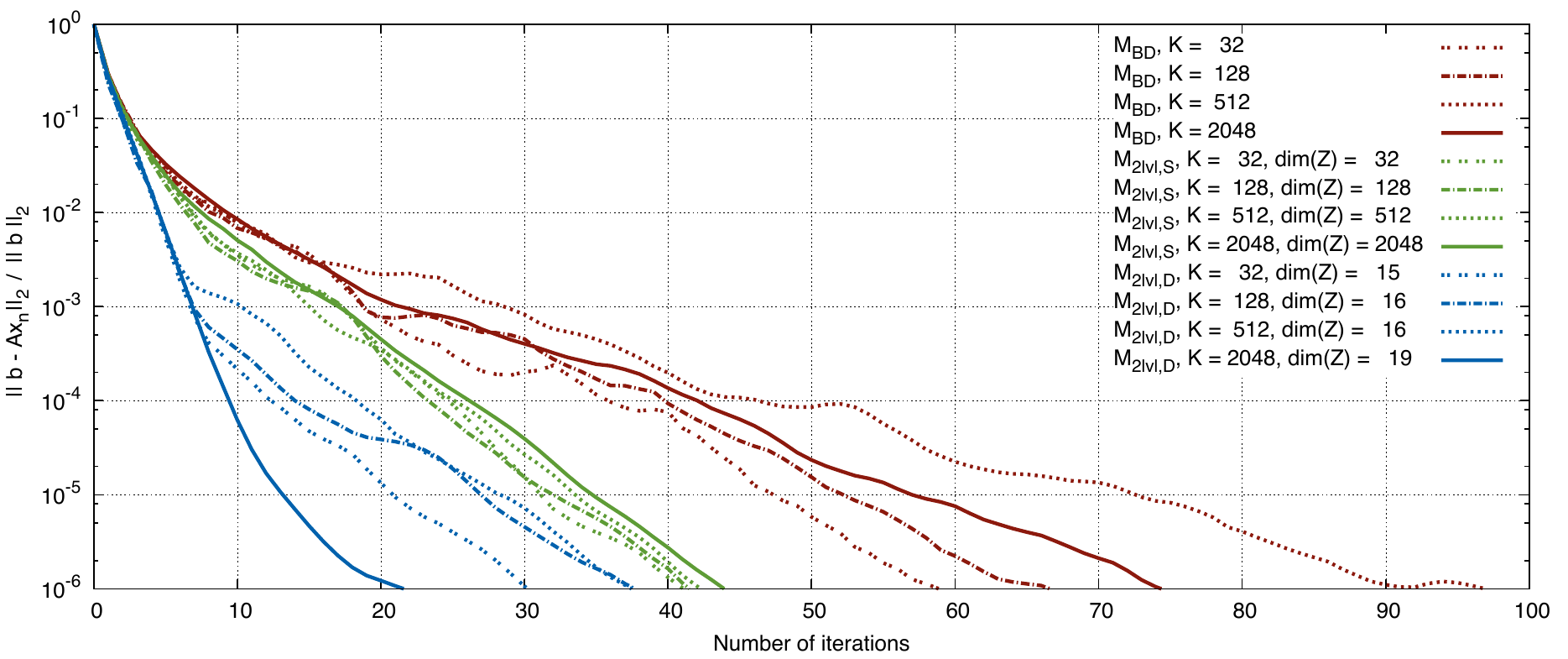}
		\end{minipage}}
		\\
		\subfloat[Polarisation case 2]{%
		\begin{minipage}[c][0.39\width]{0.84\textwidth}
		\centering
		\label{fig:convergence_pc2}\includegraphics[width=\textwidth]{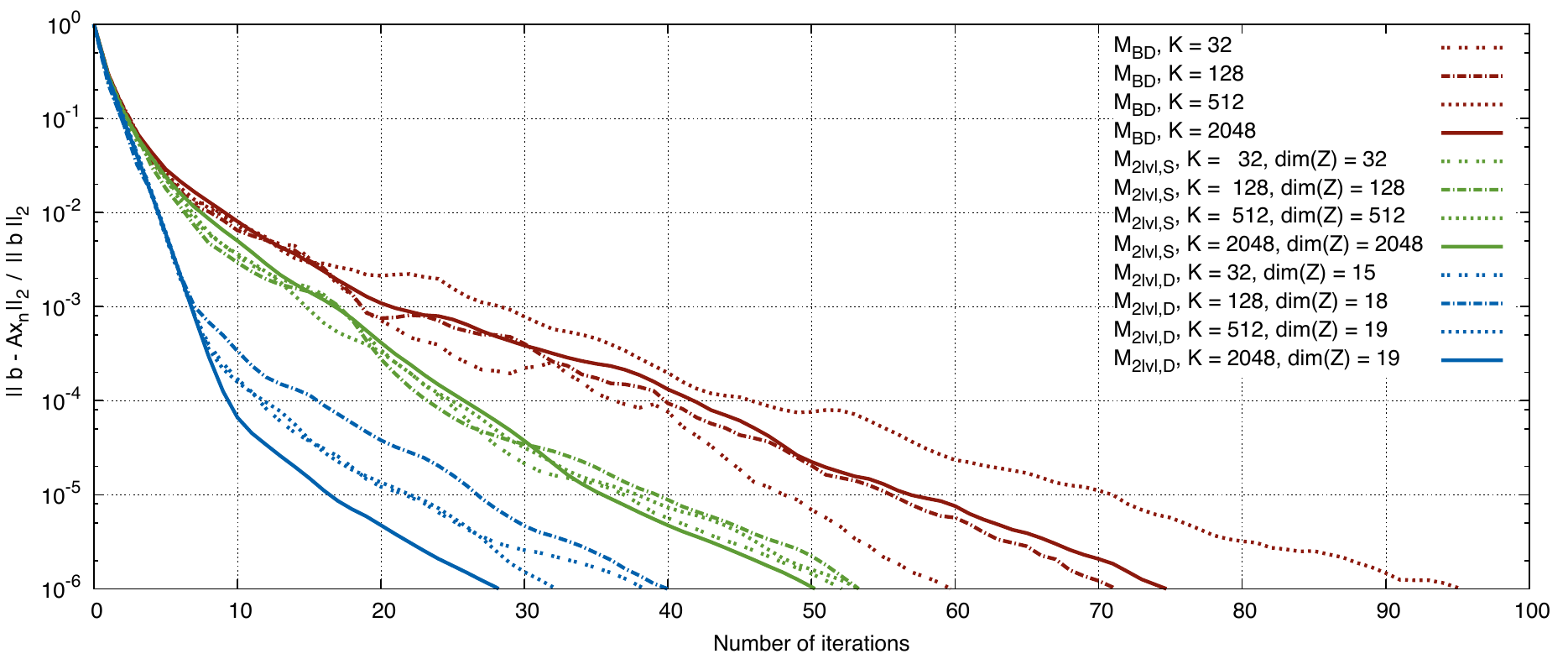}
		\end{minipage}}
		\\
		\subfloat[Polarisation case 3]{%
		\begin{minipage}[c][0.39\width]{0.84\textwidth}\centering
		\label{fig:convergence_pc3}\includegraphics[width=\textwidth]{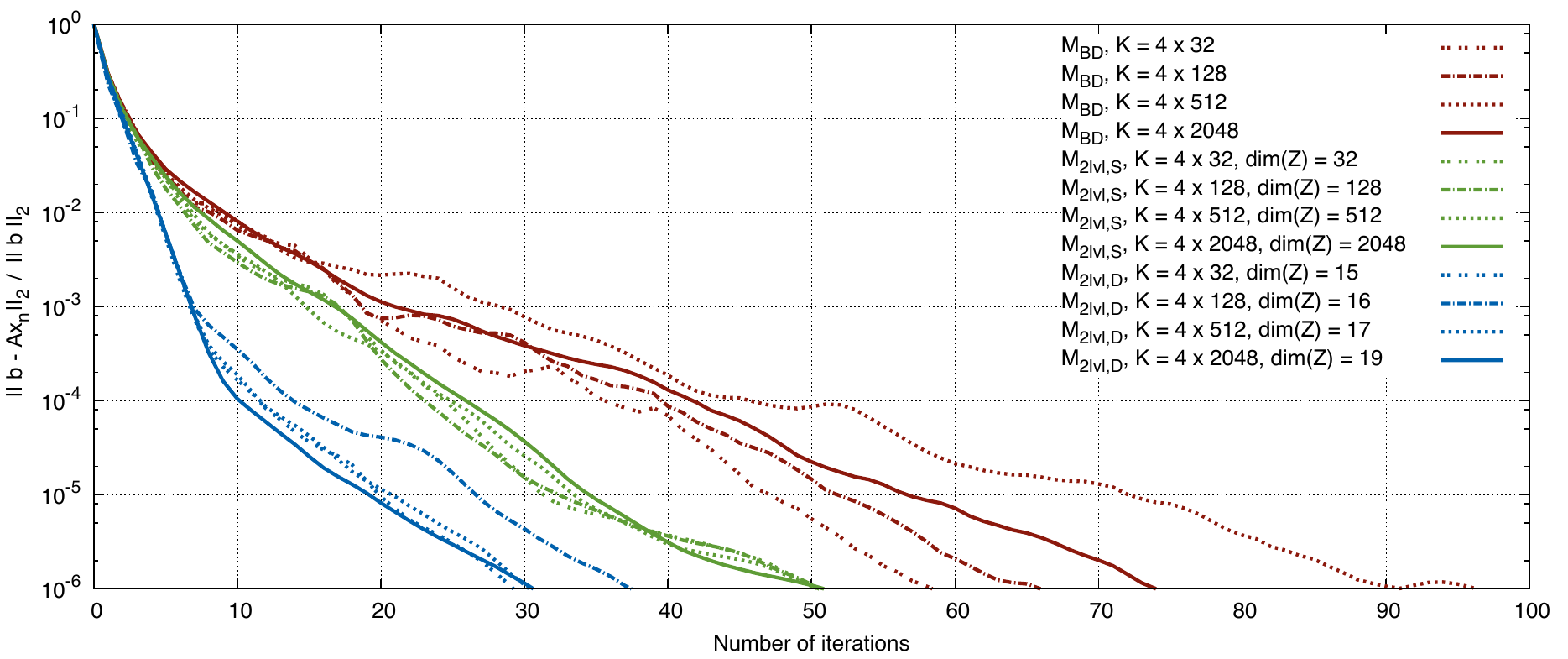}
		\end{minipage}}
		\\
		\caption{Convergence of the preconditioned CG solver applied to the simulated data, as
    described in Section~\ref{sub:performance_analysis}.  Different
    line styles correspond to different sizes of the problem, while
    different colours correspond to the three different types of
    preconditioners: red for the standard block preconditioner $\mathbf{M}_{BD}$, and green  and blue for the two-level preconditioners, $\mathbf{M}_{2lvl,S}$ and
    $\mathbf{M}_{2lvl,D}$, respectively. In the legend, $K$ corresponds to the number of stationary intervals in the time domain data and dim$(\mathbf{Z})$ denotes the number of columns in the deflation subspace operator, $\mathbf{Z}$.  For the {\em a posteriori} preconditioner, $\mathbf{M}_{(2lvl, D)}$, dim$(\mathbf{Z})$ is the number of eigenvalues, which fulfill the condition ${\rm Re}|\lambda_{i}|< \epsilon_{tol} = 0.2$.\\ \\}
\label{fig:convergence}
\end{figure*}

\begin{figure*}[!ht]
  \subfloat[Rectangular scan.]{%
	\begin{minipage}[c][0.59\width]{0.45\textwidth}
	   \centering%
	   \label{fig:def_rect}\includegraphics[width=\textwidth]{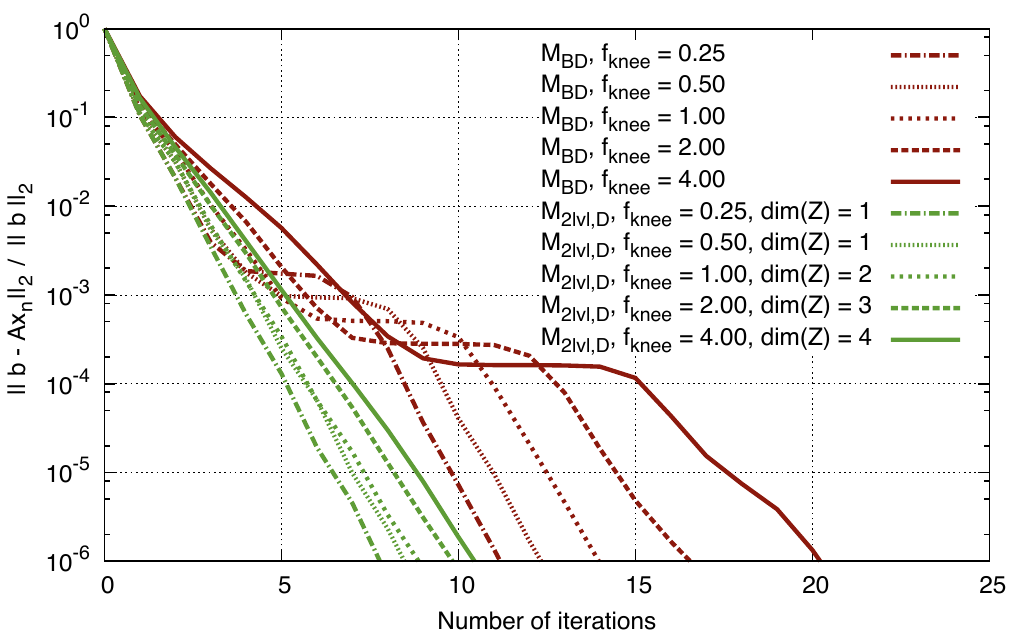}
	\end{minipage}}
	\hfill  
  \subfloat[Small circles scan.]{%
	\begin{minipage}[c][0.59\width]{0.45\textwidth}
	   \centering%
	   \label{fig:def_circ}\includegraphics[width=\textwidth]{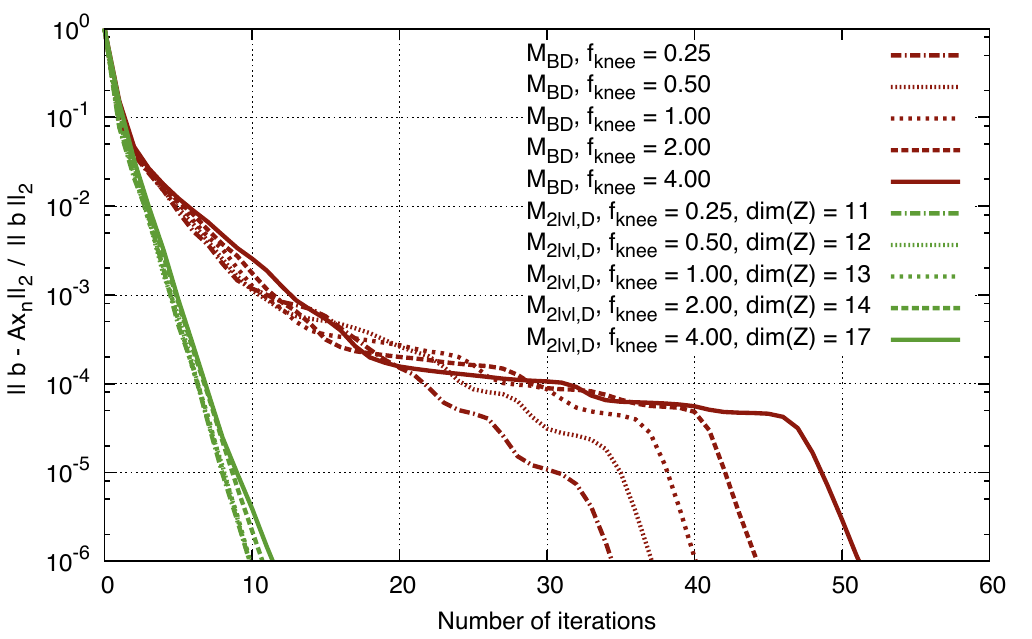}
	\end{minipage}}
\caption{Convergence of the PCG solver of the CMB map-making problem
  for two different scanning strategy. Different types of lines
  correspond to different versions of the Toeplitz matrix used in the
  simulation, as described by the parameter $f_{knee}$.  Different
  colours correspond to the two different types of preconditioners:
  red for the standard block preconditioner $\mathbf{M}_{BD}$ and
  green for the two-level preconditioner $\mathbf{M}_{2lvl,D}$
  whose coarse space $\mathbf{Z}_{D}$, is made of approximated
  eigenvectors corresponding to small eigenvalues $\lambda_{i}$ of
  $\mathbf{M}_{BD}\,\mathbf{A}$. dim$(\mathbf{Z})$ denotes
  the number of columns in the deflation subspace operator
  $\mathbf{Z}_D$ which is the number of eigenvalues
  with $({\rm Re}|\lambda_{i}|< \epsilon_{tol} =
  0.3).$}\label{fig:deflation_iterations}
\end{figure*}

Figure~\ref{fig:convergence} shows the magnitude of the relative error
of the solution defined as $(||\vec{b}-\mathbf{A}\,\vec{x}_{i}||_{2}
/ ||\vec{b}||_{2})$ as a function of the number of iterations for  the standard and two two-level preconditioners: {\em a
  priori} and {\em a posteriori}. 
 The results shown here are derived for the scanning
strategy referred to as case three in
Section~\ref{sub:scanning_strategies} 
and the different panels correspond to the three variants for
polariser dynamics, as described in Section~\ref{sub:polariser}.
In each panel, we show results for different lengths of the time domain
data, varying the number of assumed sky circles  from $32$ to $2048$. 
For cases 1 and 2 of the polariser rotation, each circle corresponds to
a stationary interval, and we alternate the value of $f_{knee}$ between
$0.5$Hz and $1$Hz for odd and even circles, respectively. For case 3, each circle
is scanned four times, each time with a different $f_{knee}$ alternating as above.
The number of stationary intervals, $K$, is marked in each panel for each of the
studied cases. For the {\em a priori} construction, $K$ defines the dimension of the deflation subspace and the number of columns of
$\mathbf{Z}_S$.
The {\em a posteriori} operator,
$\mathbf{Z}_D$, is built from approximated eigenvectors, which are
selected assuming $\epsilon_{tol} = 0.2$.  The number of these eigenvectors is
given by ${\rm dim}\,(\mathbf{Z})$, as displayed in each panel.
	
In terms of the number of iterations needed to reach some
nominal precision, assumed hereafter to be $10^{-6}$. It is apparent from the plots that the two-level preconditioners
outperform the standard one by as much as twice for the {\em
 a priori} two-level preconditoner and a factor of $3.5$ for the {\em
  a posteriori} one. 
  This confirms the theoretical expectations from Section~\ref{sec:two-level-prec}.  

Clearly, both two-level preconditioners offer important performance gains.
However, these are more modest compared to those
 found in~\citet{midas_sc12}, who reported typical
reduction in the number of iterations by as much as a factor of $6$. We attribute this to two
facts. First, the bandwidth of the Toeplitz matrices assumed in this
work was shorter than what was used in the previous work. Second, the
long-time scale noise correlations are more important in the case of
total intensity, as studied in~\citet{midas_sc12}, than in the
polarised case, as considered in this work. We elaborate on these issues
in the following sections.

\begin{figure*}[!t]
  \subfloat[With polarisation (Case 2)]{%
	\begin{minipage}[c][0.75\width]{0.45\textwidth}
	   \centering%
	   \label{fig:adapt_nic_polar}\includegraphics[width=\textwidth]{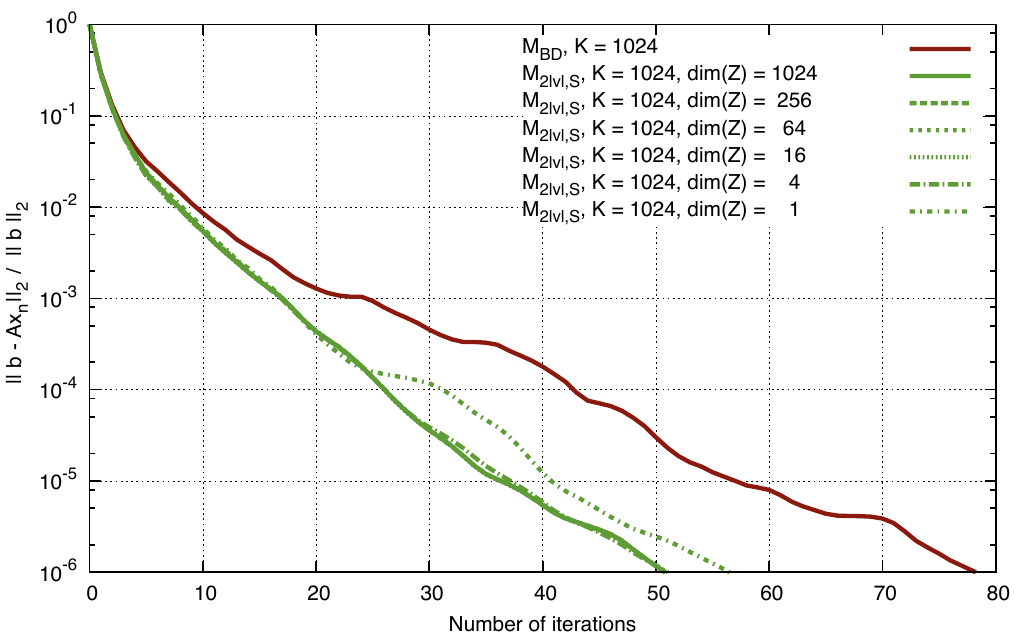}
	\end{minipage}}
	\hfill  
  \subfloat[No polarisation]{%
	\begin{minipage}[c][0.75\width]{0.45\textwidth}
	   \centering%
	   \label{fig:adapt_nic_no_polar}\includegraphics[width=\textwidth]{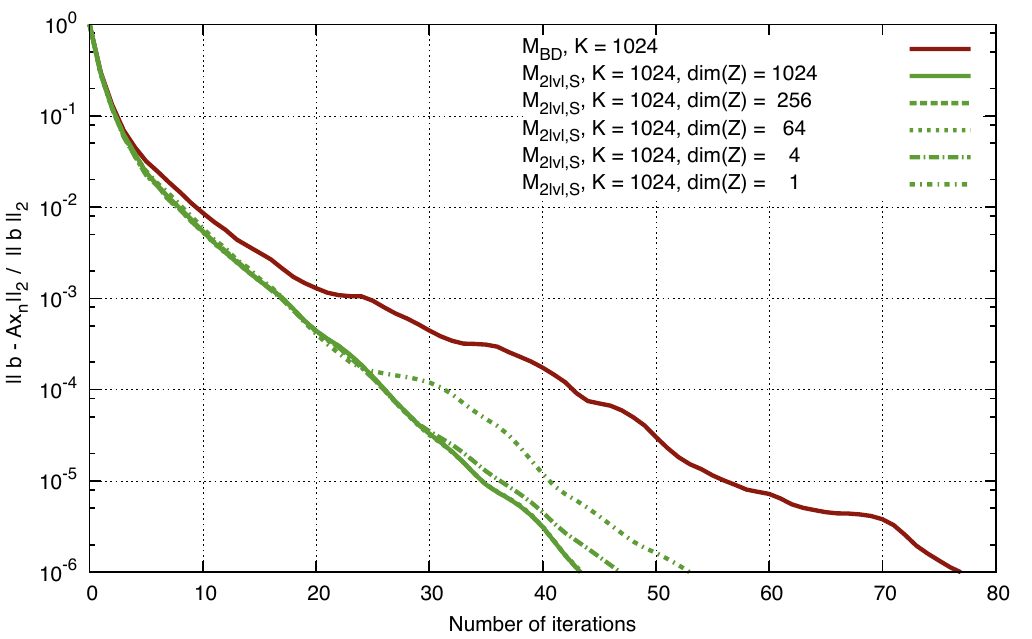}
	\end{minipage}}
\caption{Convergence of the CG solver preconditioned with the two-level {\em a priori} precondtioner for the scan made of $K = 1024$ big circles. Green lines correspond to the solutions obtained using the deflation subspace matrix, $\mathbf{Z}_S$, with a progressively lower rank, which is given by ${\rm dim}(\mathbf{Z})$ in the legend. Red curves depict  the solution with the standard preconditioner, $\mathbf{M}_{BD}$.}\label{fig:adapt_nic}
\end{figure*}

\subsubsection{Convergence rate and low frequency noise correlations}\label{sub:long_term_correlations}

We illustrate the dependence on a level of the noise correlations
in Figure~\ref{fig:deflation_iterations}, which shows the convergence
of the preconditioned CG solver for several different data sets that
correspond to different values of the parameter $f_{knee}$.  In the
cases shown in the figure, we used simulations with scanning
characteristics that correspond to cases $1$ and $2$ and, therefore,
assume only one stationary period for the entire scan as seen in Table~\ref{tab:cases}. 
 In the plot,  red curves
show the results of the runs with the block diagonal
preconditioner. It is clear that the convergence of
the block diagonal preconditioner depends very strongly
on the value of  $f_{knee}$, whenever the required convergence precision is
better than $\sim 10^{-4}$. This is around this level that the relative residuals 
reach typically a plateau  and continue their decrease only once it is over.
The length of the plateau depends on $f_{knee}$
 and so does the number of iterations 
needed to reach our nominal precision level of $10^{-6}$.

This plateau in the convergence of iterative methods is usually
attributed to the presence of small eigenvalues in the spectrum of the
system matrix, $\mathbf{M}_{BD}\,\mathbf{A}$ (see
e.g. \citet{Morgan_GMRES, kharchenko95:_eigen_trans_based_precon_for,
  Chapman96deflatedand, Gersem_2001}).  In our case, the presence of
small eigenvalues depends on the level of the noise correlations, as
does the length of the stagnation phase, as indeed observed
empirically.  As our two-level preconditioners have been designed
expressly to deal with the effects due to small eigenvalues, we expect
that they should be able to help with the convergence slow-down as
seen for the standard preconditioner. This is indeed confirmed by our
numerical results shown in Figure~\ref{fig:deflation_iterations},
where green lines show the results obtained with the two-level {\em a
  posteriori} preconditioner.

The upshot of these
considerations is that the overall performance gain expected from the
two-level preconditioners will strongly depend on the assumed noise
properties. We point out that the dimension of the deflation subspace
operator, $\mathbf{Z}_D$, used in these runs increases with an
increasing value of $f_{knee}$. This is because the number
of small eigenvalues of $\mathbf{M}_{BD}\,\mathbf{A}$ also
increases, and more columns of $\mathbf{Z}_D$ are needed to correct for
them. This manifests the adaptive character of this version of the
two-level preconditioner and should be contrasted with the {\em a
  priori} one, for which the dimension of $\mathbf{Z}_S$ would have
been kept constant in all runs and equal to $1$. This would
unavoidably result in its inferior performance. We note that we recover a gain of a factor of $5$
between the standard and two-level preconditioners for the largest value of $f_{knee}$ and
case $2$ scanning strategy, which is close to
the results reported in~\citet{midas_sc12}. \goodbreak We emphasise that the
number of small eigenvalues does not depend only on the value of
$f_{knee}$, but also on where the sky signal resides in the frequency
domain (see Appendix~\ref{app:corrLength} for an analogous argument in
a somewhat different context). For fixed $f_{knee}$ the number of small eigenvalues increases if the signal shifts towards the smaller
frequencies (as a result, for instance, of a decreasing scan
speed). Consequently, both these factors play a role in
determining what gain one can expect from applying the two-level
preconditioner.

\subsubsection{Convergence rate and the deflation space dimension}\label{sub:coarse_space_impact}

\begin{figure}[!ht] 
	   \centering
	   \includegraphics[width=0.45\textwidth]{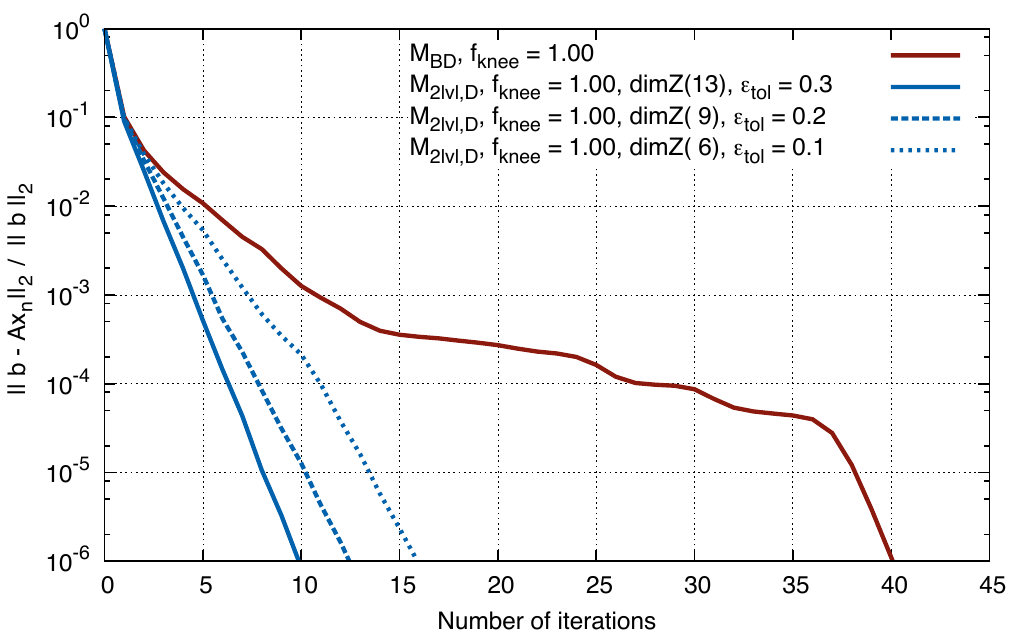}
\caption{Convergence of PCG solver preconditioned with the two-level {\em a posteriori} preconditioner computed for the small-circles scan with $K = 128$ circles and $f_{knee} = 1.0$. Blue lines correspond to the solutions obtained with the deflation subspace operator, $\mathbf{Z}_D$, with different ranks as defined by the number of approximated eigenvectors with eigenvalues smaller than $\epsilon_{tol}$, as indicated in the legend. Red curve shows the solution with the standard preconditioner, $\mathbf{M}_{BD}$. }\label{fig:adapt_def}
\end{figure}

As is evident from the discussion above, the rank of the deflation
subspace operator, $\mathbf{Z}$, is yet another parameter with a potentially crucial
impact on the performance of the two-level preconditioners. We 
could expect that increasing the operator's size has to translate
directly into a faster convergence, as it can carry information about
more peculiar eigenvalues of the system matrix.  This expectation is indeed
supported by our numerical experiments. However, we also find that the gain quickly
decreases with growing dimension of the deflation space. This is demonstrated in
Figure~\ref{fig:adapt_nic} for the {\em a priori} preconditioner and
in Figure~\ref{fig:adapt_def} for the {\em a posteriori} one.  In  an extreme case, as shown in the former figure, nearly the entire
gain from the two-level construction is already obtained assuming a one
dimensional deflation subspace. This single eigenvalue is related to the overall offset of the
recovered map, as discussed in Sect.~\ref{sub:a_priori_construction}. For fast,
well-connected scan speeds and weak noise correlations, it may
be sufficient to take care of this major small eigenvalue.  This is
indeed the case in the example shown in Figure~\ref{fig:adapt_nic}.

The dimension of the deflation subspace also has important consequences for
the computational cost of the application of the preconditioner.  Its
adopted value should be ideally a result of a trade-off
between these two effects: the expected decrease of the number of
iterations and the expected increase in the calculation time per
iteration. We discuss the performance in terms of algorithm
runtime in the next section.

\begin{figure*}[!ht]
	\centering
	\includegraphics[width=\textwidth]{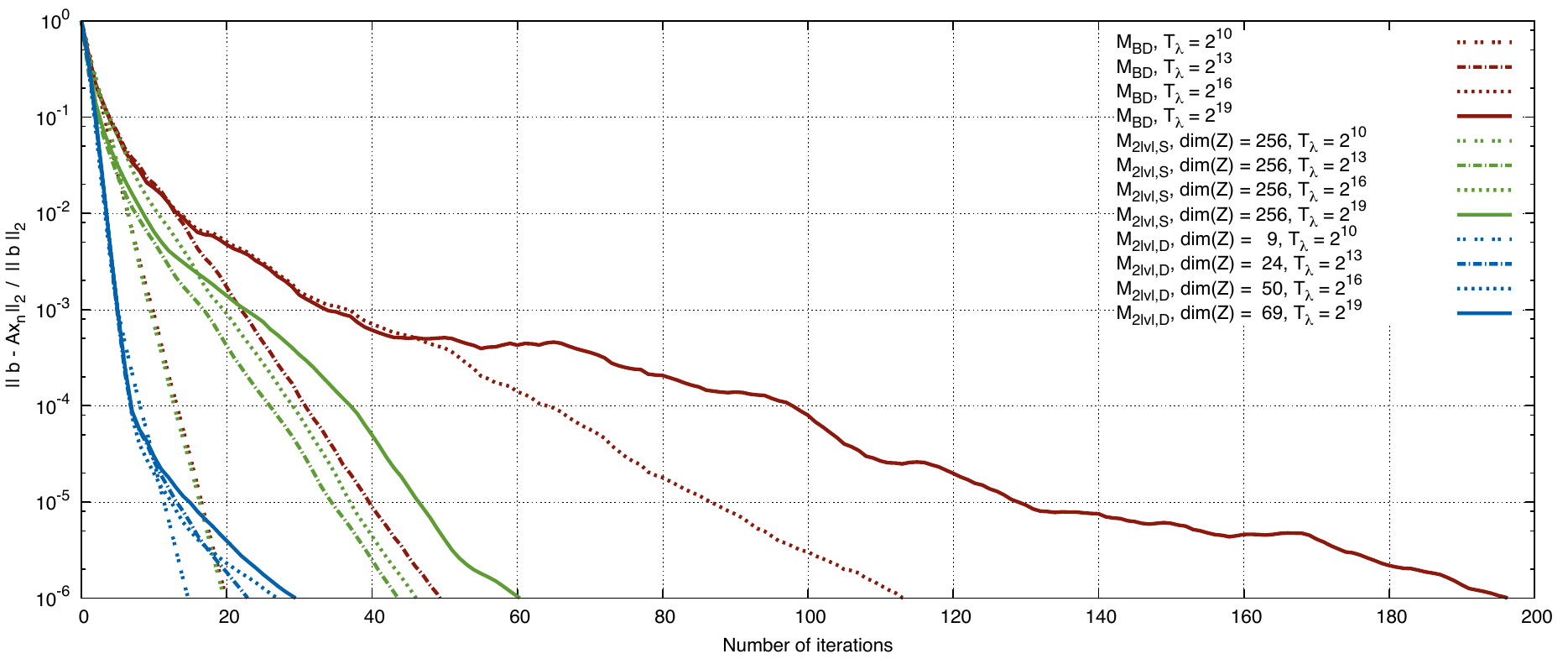}
	\caption{Convergence of PCG solver of the simulated CMB map-making problem with the same noise power spectrum but for different lengths of the imposed band limit and for the three  preconditioners considered here.}\label{fig:conv_lambda}
\end{figure*}

\subsubsection{Convergence rate and the bandwidth of the inverse noise correlation matrices }\label{sub:experiment_with_diff_band_limit} 

In the remainder of this section, we discuss the role of the Toeplitz
matrix bandwidth $\lambda_T$, which has been assumed to be fixed in
all our numerical tests presented so far with a value of
$\lambda_T = 2^{13}$. The importance of this parameter for the total
computational time needed to estimate a map is generally recognised
and well known~\citep[e.g.,][]{madmap}.  It stems from the fact that
the calculation of a product of the inverse noise matrix,
$\mathbf{N}^{-1}$, by a time-domain vector is one of the main
operations, which needs to be performed recurrently, while solving for
the map estimate. At the very best, this cost can be brought down to
${\cal O}({\cal N}_T\,\log\,\lambda_T)$, where ${\cal N}_T$ is the
length of the stationary interval, but even then it still constitutes a significant, and
often dominant, fraction of the entire run-time of the map-making
solver~\citep[][]{madmap}.  It is, therefore, desirable to select as
small a value of $\lambda_T$ as possible.

However, a value of $\lambda_T$ generally also has a direct impact 
on
the number of iterations needed to converge to the solution. This observation, which is less broadly 
recognised, is
illustrated in Figure~\ref{fig:conv_lambda}. For the standard preconditioner, we can see that changing the bandwidth can lower the
number of iteration by as much as a factor of $10$. Combined with the gain in computation time per iteration, as mentioned
above, this may seem to suggest that this is the parameter one should focus on while
optimising any map-making code performance.  We argue that though it is clearly very important, the gain in
terms of number of iterations, which can be obtained in practice from
manipulating the value of $\lambda_T$ is quite limited, and does not
supersede the gain, which can be achieved thanks to our
preconditioners.

The restriction here is due the affect a too small value of $\lambda_T$ may have on the quality of the estimated map.
 A discussion
of this issue is presented in Appendix~\ref{app:corrLength}, where we
also discuss parameters playing a role in defining the appropriate
value for $\lambda_T$. We note that such a critical value
of $\lambda_T$ is generally a result of a trade-off between acceptable loss
of precision and code performance. Moreover, determining it may not be trivial, in particular,
for complex scanning strategies.  These problems should be contrasted
with the preconditioning techniques discussed in this paper, which
attempt to speed up the calculation without any impact on the
solution.

For the cases shown in Figure~\ref{fig:conv_lambda}, the
{\em a posteriori} preconditioner delivers an order of magnitude
improvement in terms of the number of
iterations over the run based on the application of the standard
preconditioner with $\lambda_T=2^{19}$.  This is comparable to the gain achieved by simply
decreasing $\lambda_T$ by a factor of $2^9=512$.  However, the
final bandwidth, if adopted, would affect the quality of the produced
map. Indeed, in the examples studied here, the value of
$\lambda_T$ should be equal to larger than $2^{13}$ to ensure its high
precision. The two-level preconditioners yet again deliver competitive and robust performance here with a gain of a factor $2.5$ 
over the standard run with $\lambda_T =
2^{13}$ and do so without compromising the quality
of the solution. 

For the two-level preconditioners, the dependence of 
the number of iterations  on $\lambda_T$ is also much weaker than in the standard case, and, hence, the major 
driver for keeping the bandwidth
as small as possible in this case is  the
need to minimise the overall time of the computation, as we discuss in the next section.


\begin{figure*}[!t]
  \subfloat[Weak scaling.]{%
	\begin{minipage}[c][0.59\width]{0.45\textwidth}
	   \centering%
	   \label{fig:time_weak}\includegraphics[width=\textwidth]{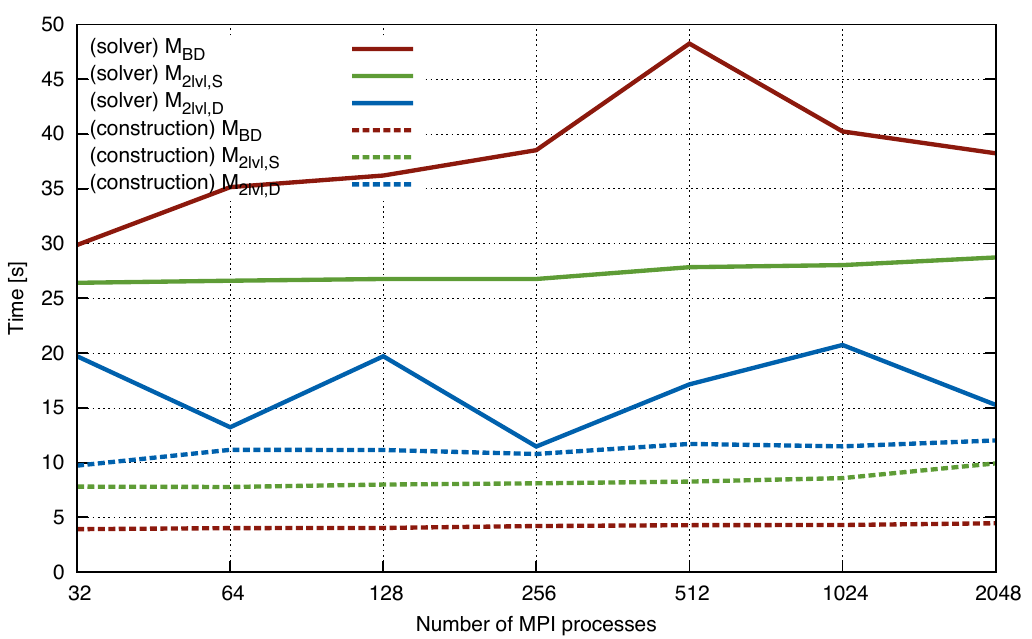}
	\end{minipage}}
	\hfill  
  \subfloat[Strong scaling.]{%
	\begin{minipage}[c][0.59\width]{0.45\textwidth}
	   \centering%
	   \label{fig:time_strong}\includegraphics[width=\textwidth]{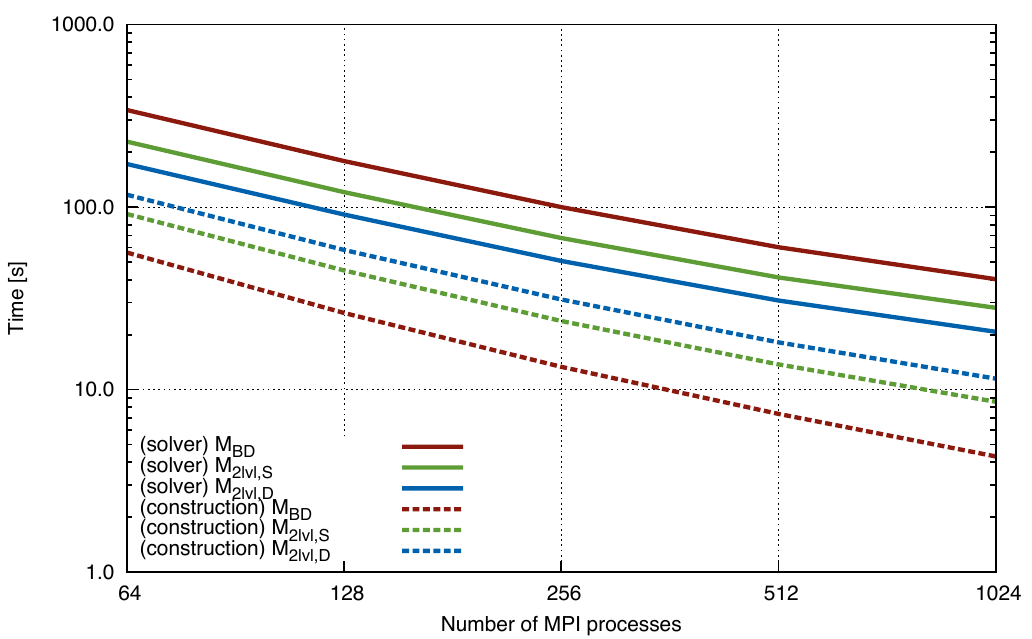}
	\end{minipage}}
\caption{Total time spent in the iterative solver for the standard, block-diagonal preconditioner, $\mathbf{M}_{BD}$, and the two variants of our new two-level preconditioner, $\mathbf{M}_{2lvl}$, plotted as a function of the number of MPI process used for the computation. 
The results of the weak scaling tests, i.e., in which problem size is increased proportionally to the number of
employed MPI processes, are shown in the left panel, while the results of the tests with a fixed size of the problem, so called strong scaling, are depicted in the right panel. 
Dashed lines show the cost of the construction of the preconditioners.}\label{fig:overall_time}
\end{figure*}

\begin{figure*}[!t]
  \subfloat[Weak scaling.]{%
	\begin{minipage}[c][0.59\width]{0.45\textwidth}
	   \centering%
	   \label{fig:speedup_weak}\includegraphics[width=\textwidth]{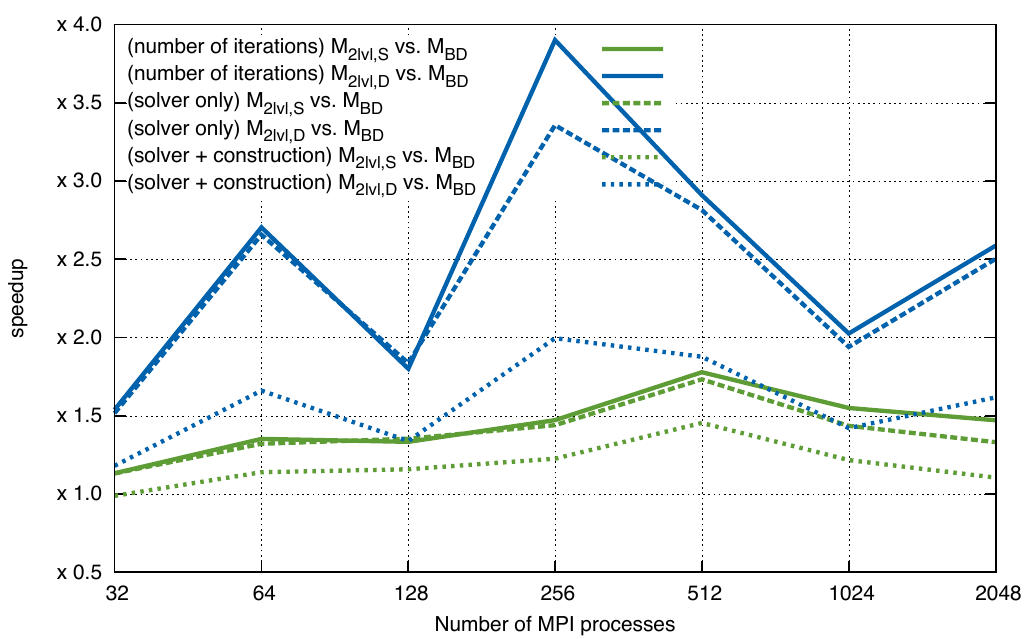}
	\end{minipage}}
	\hfill  
  \subfloat[Strong scaling.]{%
	\begin{minipage}[c][0.59\width]{0.45\textwidth}
	   \centering%
	   \label{fig:speedup_strong}\includegraphics[width=\textwidth]{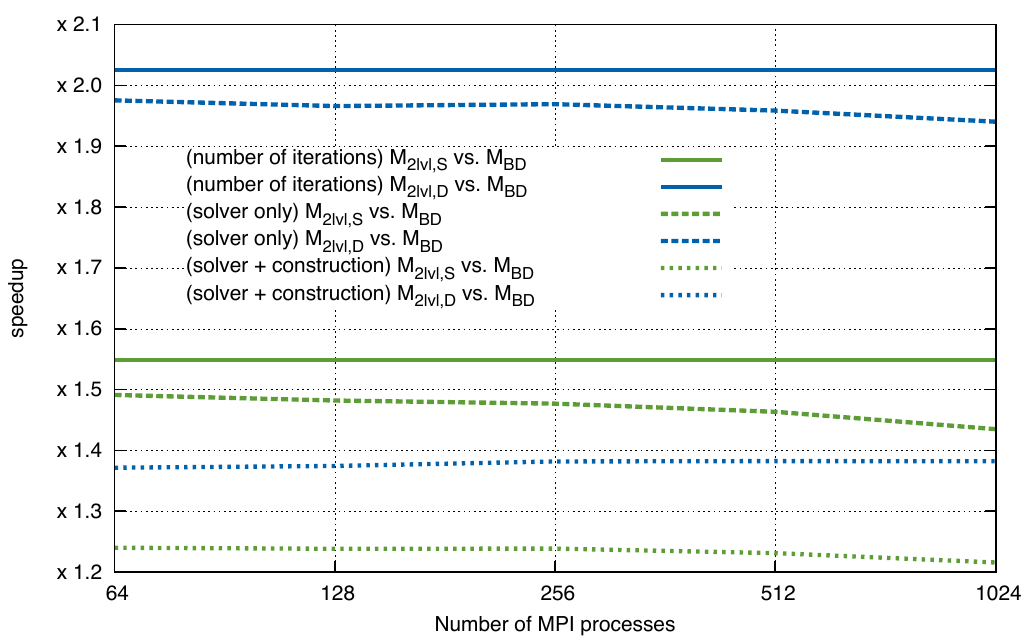}
	\end{minipage}}
\caption{Solid lines show the speedups achieved by the methods  with the two-level preconditioners, $\mathbf{M}_{2lvl, S}$ and $\mathbf{M}_{2lvl, D}$,  with respect to the traditional solver with the block diagonal preconditioner, $\mathbf{M}_{BD}$, while dotted and dashed lines show the speedups obtained on different stages of the solution.
The panels show weak (left), and strong (right), scaling, respectively.
}\label{fig:overall_speedup}
\end{figure*}

\subsection{Runtime comparison}\label{sub:performance_analysis} 

In this section, we evaluate the performance of the proposed methods in terms
of the time needed for solving the map-making problem by separately measuring
the time required by each solver to construct the
preconditioner and time spent on the iterative method.
These results are
summarised in Figures~\ref{fig:overall_time}
and~\ref{fig:overall_speedup}, showing the runtime results and achieved
speed-ups, respectively.


We emphasise that we have not performed any tuning of
the deflation subspace dimension for these runs, which could further improve 
performance of the two level  preconditioners. We
also use the standard value for $\lambda_T(=2^{13})$ and a moderate value of
$f_{knee} (=1Hz)$. As was shown earlier, these two parameters affect the
performance of the standard preconditioner significantly more than that of the two-level ones, so the timings shown here
should be seen as rather conservative.

Figure~\ref{fig:overall_time} shows the timing results. Dashed
lines depict times needed to construct the preconditioners, and
solid lines display time required by the iterative solver to
converge to the nominal tolerance of $10^{-6}$.  These timings are shown as a function
of the number of MPI processes used for the runs. The left panel shows
the so-called weak scaling, when the size of the problem
increases proportionally to the number of MPI processes.
The right panel shows the strong scaling when the problem size is
fixed.  In our weak scaling performance tests, we 
assign a single stationary interval and, therefore, a single diagonal block of $\mathbf{N}^{-1}$ to a single MPI process. Increasing the number of MPI processes concurrently increases 
the length of the data set and the number of stationary
intervals.  In the strong scaling tests, we fix the number of circles
to $1024$ and distribute the data evenly among the available
processes, by assigning multiple blocks to one MPI process. This limits
the number of processes which can be used in these tests to $1024$,
and, therefore, the number of cores to $6144$.

The results shown in Figures~\ref{fig:overall_time}
and~\ref{fig:overall_speedup} demonstrate  that the two-level preconditioners fare
better in all cases and often much better than the standard one with the {\em a posteriori} preconditioner found 
to be consistently  the best. The speedups over the standard case can be as large as $3.5 \sim 4$ and 
tend to be on average around $\sim 2.5$ for the {\em a posteriori} option and on the order of $1.5$ for 
the {\em a  priori} one. 
The two-level {\em a posteriori} preconditioner shows superior performance even when the preconditioner construction 
time is taken into account, despite its construction taking the longest out of the three options considered in this work.
The respective speedups are consequently lower and typically around $\sim 2$ for the {\em a posteriori} preconditioner. 
For the {\em a priori} preconditioner the speedup is minor but still consistently better than $1$.
Importantly, if successive linear systems with the
same system matrix but different right-hand sides need to be solved, the cost
of building the preconditioner needs to be incurred only once, and we can
eventually recover the speedups of $\sim 3.5$ and $\sim 1.5$ for the {\em a posteriori} and
{\em a priori} two-level preconditioners, respectively.

We note that the limitations on the speedups obtained here are not due to our numerical implementation
but stem from the method itself. This can be seen by comparing the speedups achieved by the code without
the precomputation time (dashed lines in Figure~\ref{fig:overall_speedup}) with that estimated solely on the basis 
of the improvement in the number of iterations needed for the convergence (0solid lines). Both these speedups
are found to be close, even though the latter estimates do not account for the extra time needed to apply the more 
complex two-level preconditioners. We conclude that there is little we could gain by optimising our
code at this time. This is supported by the discussion of the overall time breakdown between specific operations
given in Appendix~\ref{sec:parall-constr-coarse}.

The results also show very good scaling properties of the proposed two level algorithms. Indeed, 
the time
needed to solve a fixed-size problem decreases nearly inversely proportionally to the number
of employed MPI processes (strong scaling), while it remains essentially constant if the
problem size grows in unison with the number of the process (weak scaling). This potentially
shows that these algorithms eventually can be run on 
a very large number of distributed compute nodes, as required by forthcoming huge data sets
and can do so without significant performance loss.




\section{Conclusions}\label{sec:conclusions} 

We have proposed, implemented, and tested two new iterative algorithms suitable for solving the map-making problem in the context of scanning, polarisation-sensitive 
CMB experiments. 
These algorithms 
are based on the PCG method supplied with new two-level preconditioners, which are constructed either on the basis of global 
properties of the considered data set, referred to as {\em a priori} preconditioners, or with help of some specialised precomputation, {\em a posteriori} preconditioners. 
This work, therefore, generalises the considerations of~\citet{midas_sc12} in two directions. First, it considers polarisation-sensitive observations. Second,
it studies a broader class of preconditioners. 

With the help of numerical
experiments, we have demonstrated that the proposed solvers consistently lead to better performance than the standard, block-diagonal preconditioners, which define
the state-of-the art in the field. In particular, we find that the {\em a posteriori} preconditioner can decrease the number of iterations needed for convergence by as much as a 
factor of $5$, and the overall runtime by as much as a factor of $3.5$. The gains obtained with the {\em a priori} preconditioner are more modest but still interesting given
how straightforward its implementation is. 

We have studied the dependence of the performance of the proposed
preconditioners on some of the parameters defining CMB data sets.  
In particular, we have found that the performance of the
new preconditioners deteriorates only slowly when increasing the
bandwidth of the inverse noise correlations, in contrast with the
performance of standard preconditioners. This permits us to avoid
a subtle and difficult trade-off between calculation
quality and speed, which is inherently present in the latter case.
Indeed, with the two-level preconditioners we can opt for a
conservatively high value of the bandwidth, evading any danger of
compromising the quality of the estimate, while retaining the
computational efficiency.

Throughout this work, we have assumed that the time-domain noise covariance has a block-diagonal structure with
Toeplitz blocks. The proposed approaches can, however, be applied to more complex noise models as long 
as the noise weighting procedure, a prerequisite also required by the standard approach, is computationally feasible. This applies to noise models
obtained by a marginalization of unwanted systematic contributions, as discussed in \citet{Stompor02}. As these more
complex noise models will typically require more computations, the overhead due to the application of the two-level preconditioners
will be relatively smaller and their relative performance compared to the standard method better. Further gains could be expected if
these more complex noise models result in some near degeneracies of the system matrix. This may affect the performance of the standard approach
to a larger extent than that of the two-level ones.


\begin{acknowledgements}

This work has been supported in part by French National Research Agency (ANR)
through COSINUS program (project MIDAS no. ANR-09-COSI-009).
The HPC resources were provided by GENCI- [CCRT/TGCC] (grants 2011-066647 and 2012-066647) in France and 
by the NERSC in the US, which is supported
by the Office of Science of the U.S. Department of Energy under Contract No. DE-AC02-05CH11231. 
We acknowledge using the HEALPIX~\citep{HEALPix} software package and would like to thank Phil Bull and the anonymous referee for their helpful comments, which
led to improvements of this paper. We thank warmly Yousef Saad for insightful discussions.

\end{acknowledgements}

\appendix


\section{Background on Krylov subspace iterative methods}\label{app:aprox_eigenvector}

In this Appendix, we provide a brief overview of Krylov subspace
iterative methods, as well as an approach for approximating
eigenvectors of a matrix. Broader background of these techniques can
be found in, e.g., \citet{golub1996} and~\cite{saad2003iterative}.
Approximations of eigenvectors/eigenvalues of a matrix in the
context of deflation techniques are discussed in
e.g.,~\citet{Morgan_GMRES, Chapman96deflatedand}. \citet{ADDM}
gives a recent example of applications of all these techniques in the context
of preconditioners of linear systems.

There are several different classes of iterative methods for solving
the linear system of equations,
\[
\mathbf{B}\,\vec{x} = \vec{b},
\]
where $\mathbf{B}$ is a real $n \times n$ matrix, $\vec{b}$ is the right hand
side vector of dimension $n$, and $\vec{x}$ is the sought of solution vector
of dimension $n$.  Among those, Krylov subspace methods are the
most efficient and popular ones.  They belong to a more general
class of projection methods, which start from an initial guess $\mathbf{x}_0$ and,  after $m$
iterations, find an approximate solution, $\vec{x}_m$, from the affine subspace $\mathbf{x}_0
+ \mathcal{K}_{m}$, which fulfills the Petrov-Galerkin condition,
\[
\mathbf{b} - \mathbf{B} \mathbf{x}_m \perp \mathcal{L}_m,
\]
where $\mathcal{K}_{m}$ and $\mathcal{L}_m$ are some appropriately defined, method-dependent, $m$-dimensional subspaces of
$\mathbb{R}^n$.  

For the Krylov subspace methods, $\mathcal{K}_{m}$ is defined as
\begin{eqnarray}\label{eq:krylov_block_prec}
	\mathcal{K}_{m}( \mathbf{B} , \vec{r}_{0})= span \left\{\vec{r}_{0},\mathbf{B} \vec{r}_{0}, \mathbf{B}^{2}\vec{r}_{0},\ldots, \mathbf{B}^{m-1}\vec{r}_{0} \right\},
\end{eqnarray}
where $\vec{r}_0$ is the initial residual, i.e., $\vec{r}_{0} := \vec{b} -
\mathbf{B}\,\vec{x}_0$. It is referred to as the Krylov subspace of dimension $m$.
A  Krylov subspace method therefore approximates the solution as
\[
\vec{x}_m = \vec{x}_0 + p_{m-1}(\mathbf{B}) \, \vec{r}_0, 
\]
where $p_{m-1}$ is a polynomial of degree $m-1$.  
While all Krylov subspace methods are based on the same
polynomial approximation, different choices of the subspace $\mathcal{L}_m$
give rise to different variants of the method, such as a conjugate
gradient (CG), or general minimal residual method
(GMRES)~\citep{gmres86}. 

In particular, the GMRES algorithm employs the so-called Arnoldi
iterations (see algorithm~\ref{algo:arnoldi}) to build an orthogonal
basis of the Krylov subspace.  These produce the orthonormal basis,
$\mathbf{W}^{\left(m\right)}=|\vec{w}_{1}\,| \vec{w}_{2}\,| \ldots
\,|\ \vec{w}_{m}|$, of the Krylov sub-space together with a set of scalar
coefficients, $h_{ij}$, (where $ i,j = 1, \dots, m$ and $i \le j+1$)
plus an extra coefficient, $h_{m+1,m}$. The first group of the
coefficients can be arranged as a square matrix of rank $m$, called $\mathbf{H}^{\left(m\right)}$, with all elements
below the first sub-diagonal equal to $0$. A matrix with such
structure is referred to as an upper Hessenberg matrix. It can be
shown~\citep[e.g.,][]{golub1996} that there exists a fundamental
relation between all these products of the Arnoldi process, which
reads
\begin{eqnarray}\label{eq:arnoldi}
\mathbf{B}\,\mathbf{W}_{m}\,=\,\mathbf{W}_{m}\,\mathbf{H}_{m}\,+\,h_{m+1,m}\,\vec{w}_{m+1}\vec{e}^{T}_{m}.
\end{eqnarray}
Here, $\vec{e}_m$ is a unit vector with $1$ on the $m$th place.
\begin{algorithm}
\caption{Basic Arnoldi Algorithm
}
\begin{algorithmic}[1]
\REQUIRE $\vec{r}_{0}$,\,$\vec{w}_{1} = {\vec{r}_{0}}/{||\vec{r}_{0}||}$
\FOR{$j = 1 \to m$}
	\FOR{$i = 1 \to j$}
		\STATE $h_{i,j} = (\mathbf{B}\vec{w}_{j},\vec{w}_{i})$
	\ENDFOR  
	\STATE $\vec{v}_{j} = \mathbf{B}\vec{w}_{j} - \sum\limits_{i = 1}^{j}h_{i,j}\vec{w}_i$
	\STATE $h_{j+1,j} = || \vec{v}_{j} ||_{2}$ 
	\STATE $\vec{w}_{j+1} = \vec{v}_{j} / h_{j+1,j}$
\ENDFOR
\end{algorithmic}
\label{algo:arnoldi}
\end{algorithm}
The eigenpairs of the matrix $\mathbf{H}_m$ are commonly referred to
as Ritz eigenpairs.  They can be straightforwardly computed thanks to
the Hessenberg structure of the matrix and its moderate size, which is given by
the size of the Krylov space and therefore by the typical number of the
iterations needed to solve the system. This is in the CMB applications of the order of ${\cal O}(100)$.  By denoting these eigenpairs as $(\vec{v}_i,\lambda_i)$,
we can therefore write
\begin{eqnarray}
\mathbf{H}_m\,\vec{v}_i \, = \, \lambda_i \, \vec{v}_i.
\end{eqnarray}
From Eq.~\eqref{eq:arnoldi}  for every eigenvector, $\vec{v}_i$, of $\mathbf{H}_m$, we find that
\begin{eqnarray}
\mathbf{B}\,\mathbf{W}_m\,\vec{v}_i \, & = &  \, \mathbf{W}_m\,\mathbf{H}_m\,\vec{v}_i \, + \, h_{m+1, m}\,\vec{w}_{m+1}\,\vec{e}^\dagger_m\,\vec{v}_i\nonumber \\
& = & \, \lambda_i \, \mathbf{W}_m\, \vec{v}_i \, + \, h_{m+1, m}\,\vec{w}_{m+1}\,\vec{e}^\dagger_m\,\vec{v}_i.
\label{eq:approxEigen}
\end{eqnarray}
Consequently, if $h_{m+1, m} = 0$, then every eigenvector of $\mathbf{H}_m$ defines an eigenvector of $\mathbf{B}$ given by $\vec{y}_i := \mathbf{W}_m\,\vec{v}_i$,  both of which have the same corresponding eigenvalue, $\lambda_i$. 

If $h_{m+1, m}$ is not zero, but is small, as it is usually the case if the solver is converged with sufficient precision. The pairs $(\vec{y}_i, \lambda_i)$ then provide a useful approximation of the $m$ true eigenvalues of $\mathbf{B}$, and the magnitude of the last term on the right hand side of Eq.~\eqref{eq:approxEigen} serves as an estimate of the involved error. 

In our construction of the {\em a posteriori} preconditioner, we take $\mathbf{B} = \mathbf{M}_{BD}\mathbf{A}$ and apply the algorithm described above to calculate the set  of the vectors, $\vec{y}_i$, from which we  subsequently select the vectors used to build the deflation subspace operator, $\mathbf{Z}_D$. 

Though the specific approach described here relies on the GMRES solver, we note that a similar construction can be  performed using the CG technique as elaborated on, for instance, in~\cite{ACG}.


\section{Alternative construction of a two-level preconditioner}\label{sec:alt_prec}

A two-level preconditioner can be constructed in a different way than what is proposed
in Section~\ref{sec:two-level-prec}. One potentially promising alternative, which stands out from a theoretical point of view, can be derived from the 'Adapted Deflation Variant 2' method (A-DEF2) of \citet{TANG:2009} that renders the following expression,
\begin{eqnarray} \label{eq:ADEF2}
\mathbf{M}_{2lvl,}^{(alt)} := \mathbf{R^T}\, \mathbf{M}_{BD}\,+\,\mathbf{Z}\,\mathbf{E}^{-1}\mathbf{Z}^{T}.
\end{eqnarray}
This preconditioner, like the one proposed earlier, is neither symmetric nor positive
definite~\citep[][]{TANG:2009}. However, it can be more robust than the choice studied in this work in some applications. This is, because
it can be shown that there exists
an initial solution for the iterative solver in exact arithmetic such that  $\mathbf{M}_{2lvl,}^{(alt)}$ is equivalent to an SPD preconditioner.
while this is not the case for $\mathbf{M}_{2lvl}$. This is an important difference, as the
convergence of a conjugate gradient algorithm is only guaranteed for the SPD matrices.
For our choice of the two-level preconditioner, the convergence has to be tested experimentally
case-by-case. 
Nevertheless,  once this turns out to be so, as in the map-making problem studied  in this work,
the convergence rates of both these two-level constructions, are expected to be similar for the same
choice of the deflation space matrix, $\mathbf{Z}$. This is because
the spectra of $\mathbf{M}_{2lvl} \mathbf{A}$ and $\mathbf{M}_{2lvl}^{(alt)} \mathbf{A}$ 
can be shown~\citep[Theorem 3.1 of][]{Vui06NT} to be
identical.  In particular, this preconditioner shifts the 
small eigenvalues of $\mathbf{A}$ to $1$, that are set to zero  in $\mathbf{M}_{BD}\,\mathbf{R}\, \mathbf{A}$ by the deflation operator,
as does so our standard two-level preconditioner, what results in a similar clustering pattern of
the eigenvalues of the preconditioned system matrix in both these cases.

While we have experimentally confirmed all these theoretical expectations in the context of the map-making
problem, we have also found that this latter construction has a higher computational
cost and is therefore disfavoured for our application. Nevertheless, it still may provide an alternative whenever
convergence problems arise.


\section{Implementation of the two-level preconditioner}\label{sec:parall-constr-coarse}

In this Appendix, we describe major features of the parallel implementation of our two-level preconditioner.
For clarity, we comment explicitly only on the memory-distributed parallel layer of the code, assuming that one MPI process 
is assigned to every allocated processor, even though, as mentioned elsewhere, the code attempts to also capitalise on the shared-memory capabilities whenever possible.

In our code, we implement the data layout scheme of~\cite{madmap}, which has been designed to facilitate time-domain
operations and to keep communication volume low. We therefore distribute all  time domain objects by dividing all stationary intervals
into disjoint subsets and assigning data corresponding to each subset to one process. The distribution of the pointing matrix, $\mathbf{P}$,
and the inverse noise correlations, $\mathbf{N}^{-1}$, follows that of the time-ordered data with each of these objects first being divided into blocks
relevant for each stationary interval and then assigned to a corresponding process. We denote the data stored by  process $j$ as,
$\mathbf{d}_j$, $\mathbf{P}_j$, and $\mathbf{N}^{-1}_j$.

We can now define a subset of pixels, ${\cal P}_j$, as observed within
the time period assigned to process $j$. This generally is a
small subset of all observed pixels, ${\cal P}\ (\, \equiv \bigcup_j\,{\cal
  P}_j)$. Moreover, the subsets assigned to different processes may
but do not have to be disjoint, that is ${\cal P}_i \cap {\cal P}_j \ne
\emptyset$, as any given pixel may have been and, indeed, ideally has been
observed many times over the full observation.  These
subsets, ${\cal P}_j$, define the distribution of all pixel domain
objects, such as maps, the standard preconditioner, or coarse space projection in the case of the two level preconditioners
operators. The downside of such a distribution is that all these
objects are potentially distributed with overlaps. The upside is that
this restricts the communication volume.

As elaborated in~\cite{madmap}, this data distribution is particularly
efficient when calculating matrix-vector products of the matrix
$\mathbf{A}\, =\, \mathbf{P}^T\,\mathbf{N}^{-1}\,\mathbf{P}$ by a
pixel-domain vector $\vec{x}$. This is the fundamental operation
performed by any iterative map-making solver. In particular, it limits
the need for interprocess communication to a single instance, which is
an application of the de-projection operator $\mathbf{P}^T$ to a
vector of the time-ordered data.  A global reduce operation is
unavoidable to combine partial results,
$\mathbf{P}_j\,\mathbf{d}_j$, which are pre-calculated by each of the involved
processors.  This operation can be performed by calling {\tt
  MPI\_AllReduce()}~\citep{madmap}, and this is the approach
implemented here.  We note, however, that recently more efficient
solutions have been proposed~\citep{madmap_new, cargemel13}, which
scale better with the growing number of processors, and could further
improve our runtime results.

In the case of the two-level preconditioners, we also need to perform multiple
dot-products of pixel-domain objects in addition to applying
the matrix $\mathbf{A}$ to a vector. As these are distributed, this
does necessarily involve a communication of the same type, as used in applying
the deprojection operator above. Special attention has to be paid
here to the data overlaps. In the case of the dot-products if left unaccounted, overcounting contributions from
pixels, which are shared by more processors will occur. To avoid that, we
precompute the frequency with which each pixel appear on different MPI
processes and use it to weight their contribution to the final result
of the dot-product. The calculation of the frequency requires one
extra global reduce operations, which can be performed on the onset.
Once accomplished, its result is stored in the memory of each process, as 
distributed as all other pixel domain objects.

We can now describe steps involved first in constructing and then 
applying the two-level preconditioner.  
In principle, the construction needs to be performed only once, 
while the actual application has to be done at every iteration. However, 
due to the memory constraints,
the preconditioner can not be constructed explicitly and, therefore, only some of the
steps involved in the construction can be done ahead of the time, while the others will
need to be performed on every iteration, whenever the preconditioner needs to be applied.
The challenge is to find the right balance between extra memory overhead to store
pre-computation products and extra computations, which have to be 
performed repetitively.
\begin{description}
\item[{\bf Constructing the preconditioner}]{ -- we precompute two
  objects: $\mathbf{A}\mathbf{Z}$ and $\mathbf{E}$ and store them in
  memory throughout the iterative process. The latter object is stored
  as a two factor matrices computed via
  the LU factorization to facilitate its application
  later.  The involved calculations are implemented as follows,\\
\begin{enumerate}
\item $\mathbf{A}\mathbf{Z}$: We compute it by applying the system matrix,
  $\mathbf{A}$, to each column of $\mathbf{Z}$ separately, using the standard
  map-making technique outlined above and in
  Section~\ref{sub:the_maximum_likelihood_solution}. This can be time
  consuming if the number of columns, which is equal to the dimension of the
  coarse space, $r$, is large. This is, however, typically the case only
  for the {\em a priori} preconditioner, and then the computational
  load can be decreased efficiently by capitalising explicitly on the
  fact that $\mathbf{Z}$ is very sparse. The result, $\mathbf{A}\mathbf{Z}$, is
  precomputed once in the code and stored in the distributed memory.  
  Though this can be large as it amounts to storing
  multiple map-like vectors, it is typically merely a fraction of the
  volume of all the time domain data required for computation, and it leads
   to significant savings in  the operation count.\\
\item $\mathbf{E} \, = \, \mathbf{Z}^T\,\mathbf{A}\,\mathbf{Z}$: This calculation capitalises on $\mathbf{A}\mathbf{Z}$ which was
  precomputed earlier, and therefore involves only a series of weighted
  dot-products.  Once computed, the matrix $\mathbf{E}$ is decomposed
  using the LU factorisation and stored in the memory of each
  process. This is clearly superfluous. For the coarse space
  dimension considered in this work, the memory overhead is minor,
  while this choice helps to keep the code simpler.\\
\end{enumerate}
}
\item[{\bf Application of the preconditioner}]{ - on each iteration of the iterative solver, we need to apply our preconditioner to some pixel-domain vector. This involves the following operations: \\
\begin{enumerate}
\item $\mathbf{Z}^T\,\vec{x}$: For any arbitrary, pixel-domain vector, $\vec{x}$, this calculation is straightforward by using the weighted dot-product procedure that is applied to each column of $\mathbf{Z}$. As explained above, this calculation involves global interprocessor communication.\\
\item $\mathbf{E}^{-1}\,\vec{v}$: For any arbitrary, coarse space vector, $\vec{v}$, this is done by solving on each process a linear system of equation given by $\mathbf{E}\,\vec{v}_{out}=\,\,\vec{v}$. This is quick, as $\mathbf{E}$ is already precomputed and suitable decomposed using LU decomposition. This operation does not require any communication.\\
\item $\mathbf{Z}\,\vec{v}$ and  $(\mathbf{A}\mathbf{Z})\,\vec{v}$: This is done explicitly locally by each process as it does not require any communication. The second operation uses the result of the precomputation.\\
\item $\mathbf{M}_{BD}\,\vec{x}$: This is performed directly. No communication is required.
\end{enumerate}
}
\end{description}

Thus far, we have assumed that the coarse space projection matrix
$\mathbf{Z}$ is given explicitly. In
the case of the {\em a posteriori} preconditioner, it is
computed as described in
Appendix~\ref{app:aprox_eigenvector}. For the {\em a priori} case, the
 elements of $\mathbf{Z}$ reflect the frequencies with which
a given pixel appears in all the stationary intervals. Consequently,
the relevant computation is analogous to the one computing the weights 
for the weighted dot products, explained above.

From the memory load perspective, the two-level preconditioner requires extra memory to store the results of the precomputation step. However, for typical cases of interest, the time domain objects, which include the pointing matrix and the inverse noise correlations, keep on dominating  the memory requirements.

\begin{figure}[!t]
	\subfloat[Weak scaling.]{%
		\begin{minipage}[c][0.69\width]{0.45\textwidth}
			\centering%
		  \label{fig:1it_time_weak}\includegraphics[width=\textwidth]{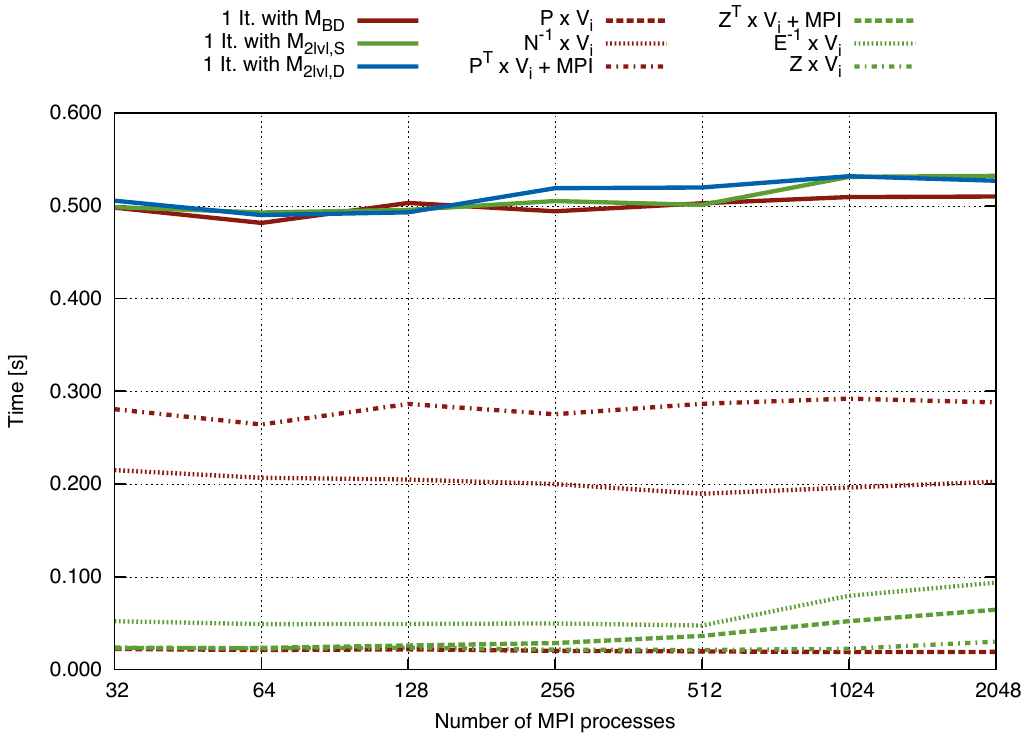}
		\end{minipage}}
		\\
	\subfloat[Strong scaling.]{%
		\begin{minipage}[c][0.69\width]{0.45\textwidth}
		   \centering%
		   \label{fig:1it_time_strong}\includegraphics[width=\textwidth]{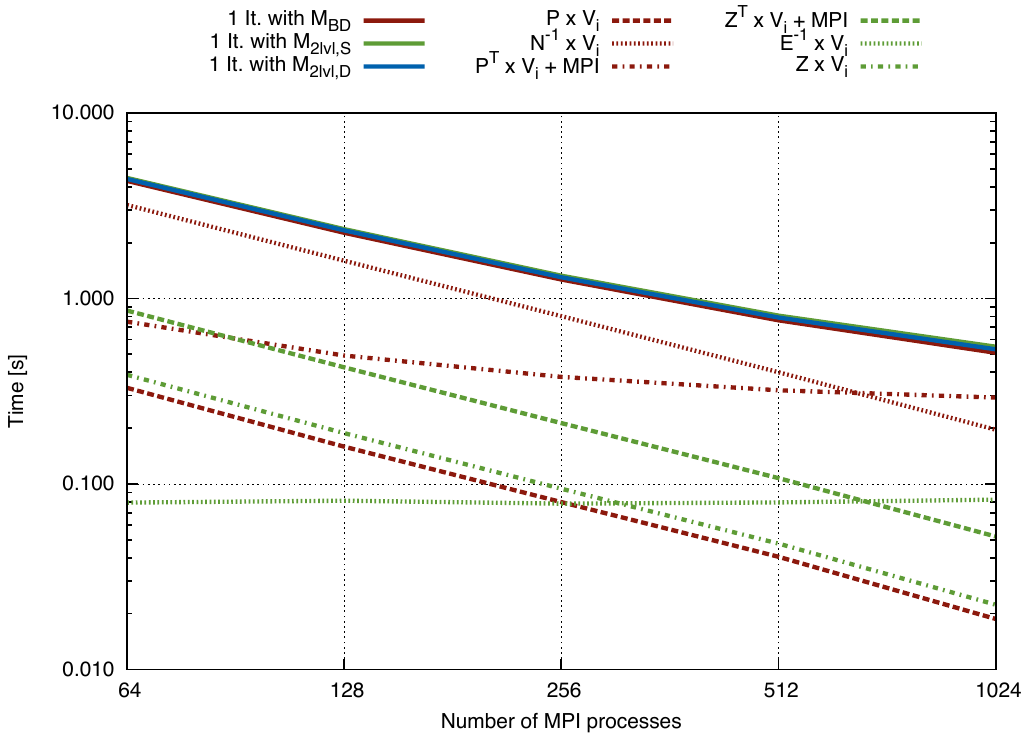}
		\end{minipage}}
\caption{Average time per iteration and its breakdown between different operations shown as a function of the number of MPI processes. The former is shown for all three preconditioners with solid lines of different colours. The latter is shown for the {\em a priori} two-level preconsitioner only. However, the time breakdown for the {\em a posteriori}
preconditioner has been found to be essentially the same in this specific experiment.
}\label{fig:it_time_general}
\end{figure}

The total time breakdown between the main steps of one iteration of the
preconditioned system is shown in Figure~\ref{fig:it_time_general}.  The $60\%$ of the overall time is spent in the
depointing operation, $\mathbf{P}^{t}_{j}\vec{x}_{j}$, which requires
communication, which is implemented in our code using a single global {\tt
  MPI\_Allreduce}.  The second most expensive operation is the
multiplication of the Toeplitz matrices by a vector, which is implemented with
help of a parallel (multithreaded) FFT algorithm.  Finally solving the
small factorized linear system to compute $\mathbf{E}^{-1}\vec{v}$ has an almost negligible impact on the
total time but becomes progressively more important for the large test cases.  However, the computations overall in the range
of considered problem sizes scale well with the growing number of processors. 
Moreover, the time per iteration of the
CG solver preconditioned by our two level preconditioner is almost the
same as the time per iteration of the CG that is preconditioned by the
standard preconditioner, as the solid red, green, and blue lines in
Figure~\ref{fig:it_time_general} nearly overlap.  This also explains the
good performance of our preconditioner, which requires a smaller number of iterations and, hence, a smaller overall time to solution.

In summary, our current implementation of the map-making software makes
a convincing case for the superior overall efficiency of the new
preconditioners proposed and studied in this work.  Though additional
optimisations can be implemented, as, for instance, included in the
MADmap code~\citep[][]{madmap}, these are not expected to change our
conclusions in any major way as they are largely derived from
comparisons of the relative performance as the major optimisations
would affect the performance of both preconditioners to a similar
extent.  Consequently, as long as the cost of solving the system with
the factorized matrix $\mathbf{E}$ and operations with $\mathbf{Z}$ remain
subdominant with respect to the remaining steps of algorithm, the
speedup factors measured in our experiments should be recovered. 


\section{Role and impact of the assumed noise correlation length}\label{app:corrLength}

\begin{figure*}[!t] 
	   \centering
	   \includegraphics[width=0.9\textwidth]{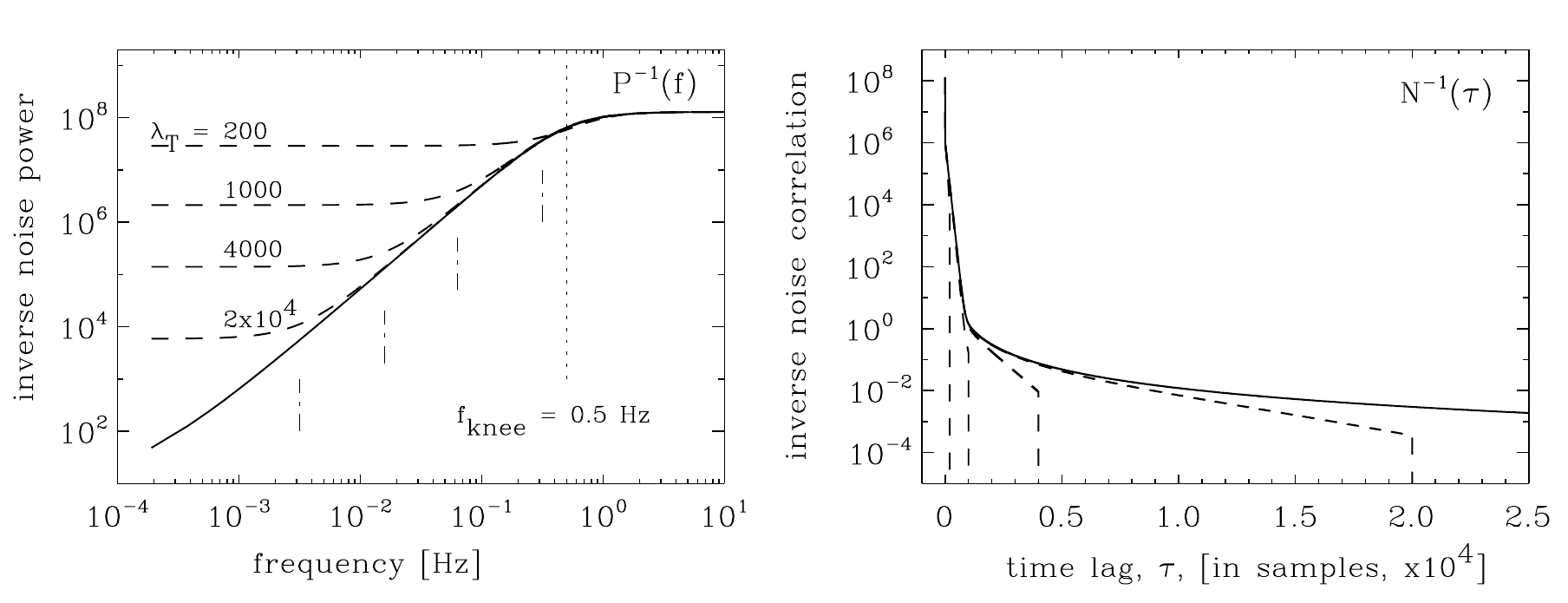}
\caption{Effects of imposing band-diagonality on the inverse noise correlation matrix, $\mathbf{N}^{-1}$, right panel, and its consequences for the inverse noise power spectrum, $P(f)$, left panel. The apodization kernel used here is defined in equation~\eqref{eq:apodDef}. The right panel shows a first row of the matrix, which is assumed to be Toeplitz. In both panels, solid lines show the result derived for the noise spectrum as in equation~\eqref{eq:fknee} without applying any apodization. Dashed lines 
show the results when the apodization is applied with the kernel length, $\lambda_T$, assuming values, $10^4, 2000, 500$ and $100$, as shown bottom-up, in the left panel, and right-to-left in the right panel. Dash-dot-dash lines in the left panel show  the characteristic frequency at which the apodized spectrum
deflects from the original one for each apodization length. This frequency is given by $f_{defl} = 1/\pi\lambda_T t_{samp}$.
}\label{fig:corrPlot}
\end{figure*}

In the presence of the off-diagonal correlations and assuming the noise correlation matrix that describes the piece-wise stationary noise present in the input time-ordered data, it is often convenient to think of the map-making problem in the Fourier rather than time domain. In this case, the data are represented as complex amplitudes of the Fourier modes as
\begin{eqnarray}
\widetilde{\vec{d}} := \mathbf{F}\,\vec{d},
\end{eqnarray}
where $\mathbf{F}$ is a block diagonal matrix with the blocks corresponding to the Fourier operators that acts independently on each stationary interval. Consequently, the sizes of the blocks
reflect those of the inverse noise correlation matrix, $\mathbf{N}^{-1}$, in equation~\eqref{eq:N_def}. Moreover, given the Toeplitz character of the blocks of the latter, we have
\begin{eqnarray}
\widetilde{\mathbf{N}}^{-1} & := & \mathbf{F}^T\, \mathbf{N}^{-1}\,\mathbf{F} \simeq  \\
&\simeq&
\begin{pmatrix}
{\displaystyle {\rm diag} \,P_0^{-1}\left(f\right)} & {\displaystyle 0} & {\displaystyle \dots} &  {\displaystyle 0}\\
{\displaystyle 0} & {\displaystyle {\rm diag} \,P_1^{-1}\left(f\right)}  & {\displaystyle \vdots} &  {\displaystyle \vdots}\\
{\displaystyle \vdots} & {\displaystyle \dots} & {\displaystyle \ddots} & {\displaystyle 0}\\
{\displaystyle 0} & {\displaystyle \dots} & {\displaystyle \dots} &  {\displaystyle {\rm diag}\, P_{K-1}^{-1}\left(f\right)}
\end{pmatrix},
 \nonumber
\end{eqnarray}
and therefore, the noise matrices in this representation  are essentially diagonal with the diagonal given by an inverse noise power spectra for each of the stationary intervals. These can be typically parametrized as in equation~\eqref{eq:fknee}.

The map-making process in the Fourier domain can be written as in equation~\eqref{eq:glsSol},
\begin{eqnarray}
\vec{m} = \left[(\mathbf{F\,P})^{\,t}\,\widetilde{\mathbf{N}}^{-1}\,\mathbf{F\,P}\right)^{-1}\,\mathbf{F\, P}^{\,t}\,\mathbf{\widetilde N}^{-1}\,\widetilde{\vec{d}},
\label{eq:glsSolFourier}
\end{eqnarray}
and comes down to weighting of the Fourier amplitudes that represents the data. $\widetilde{\vec{d}}$, by respective amplitudes of the noise power spectra, $P^{-1}_j(f)$. These are subsequently projected to the pixel domain by the projection operator, $(\mathbf{F\,P})^{\,t}$, and one corrected for by the weight matrix, $\left[(\mathbf{F\,P})^{\,t}\,\widetilde{\mathbf{N}}^{-1}\,\mathbf{F\,P}\right)^{-1}$. We note that this is analogous to what happens in the pixel domain if the time-domain noise is uncorrelated but potentially inhomogeneous. The first step of the map-making process  is then a simple noise-weighted co-addition of the data in the pixels on the sky. We point out that the resulting map will still be unbiased, but potentially noisier than necessary in both cases, if our assumptions about the time- or frequency- domain weights are wrong.

For each Toeplitz block of the full inverse noise correlation matrix, the corresponding inverse noise power spectrum can be calculated by computing a Fourier transform of one of its rows. 
For the noise models in equation~\eqref{eq:fknee}, these Toeplitz blocks are not generally band-diagonal. This is because the inverse power spectrum, as shown with a solid line in Figure~\ref{fig:corrPlot}, is never flat even at the lowest considered frequencies. If band-diagonality is desired, it would have to be therefore imposed.  This corresponds to apodizing the row of the Toeplitz matrix with an appropriately chosen apodization window. The respective inverse noise power spectrum after the apodization is then given by a convolution of the Fourier transform of
 the window and the initial inverse power spectrum.
 In the time domain, the apodization window, which is required for this task, has to fall off quickly at some required correlation length, $\lambda_T$, which defines the effective bandwidth. Its Fourier representation also does so but at the scale given by $f_{defl} \equiv 1/(\pi \lambda_T\,t_{samp})$, where $t_{samp}$ stands for the sampling rate, as in equation~\eqref{eq:fknee}. The convolution of this spectrum with the initial one therefore flattens the noise spectrum at the low frequency end with flattening that extends up to $f_{defl}$. This is illustrated in Figure~\ref{fig:corrPlot}, where we used a Gaussian window for the apodization, 
 \begin{eqnarray}
 {\cal G}( k) = 
 \left\{
 \begin{array}{l l}\medskip
{\displaystyle  \exp 2\,\left(\frac{k}{\lambda_T}\right)^2, } & {\displaystyle k \le \lambda_T;}\\
{\displaystyle 0,} & {\displaystyle {\rm otherwise.}}
\end{array}
\right.
 \label{eq:apodDef}
\end{eqnarray}
Consequently, imposing the band-diagonality modifiies the inverse noise power spectra and, therefore, weights, which are used in the map-making process, and, therefore,  generally leads to suboptimal maps. How big is the loss and, in particular, whether it is acceptable, depends on how the sky signal is distributed in the Fourier domain.  
If the sky signal resides only\footnote{We note that there is always some part of the sky signal residing at the zero frequency however, this signal is already lost whenever the time-domain noise has $1/f$-like behaviour at low frequencies independent of the choice of the bandwidth.} in the frequency above some threshold, $\simgt f_{sig}$, there is no loss of precision as long as $f_{sig} \simgt f_{defl}$ and, therefore, as long as $\lambda_T \simgt 1/(\pi f_{sig}\,t_{samp})$. In practice, such circumstances are realised only for some periodic scanning strategies~\citep[e.g.,][]{StomporWhite2004} and, more commonly, at least some part of the sky signal is present at arbitrary low, though non-zero, frequencies. In such cases, the magnitude of the precision loss clearly depends on the properties of the low frequency noise, such as its knee frequency and slope with the effect
becoming larger for larger values of $f_{knee}$ and steeper noise spectra. For given noise properties and a scan strategy, if the loss is found unacceptably large, 
it can be mitigated by appropriately increasing the bandwidth width, as is indeed the standard rule of thumb used in the map-making community. However, the extra amount of the bandwidth, which is needed to ensure some required precision, depends on the sky signal distribution in the frequency domain.

We also note that $1/(\pi t_{samp} f_{knee})$ defines a maximal  bandwidth, which is still merely equivalent to uniform weighting in the frequency domain and white noise weighting in the time domain. Therefore, only by adopting a larger bandwidth, we can obtain any improvement over the map produced with white noise weighting that is when all Toeplitz blocks are assumed proportional to unity.

The purpose of extending the correlation length is not so much to help to include the constraints coming from the very low frequency modes on the sky signal, as these are 
very noisy to start with and therefore, can anyway provide only weak constraints, but to ensure that those modes are properly weighted down, so the noise they contain does not overwhelm the constraints from higher frequencies where the noise per frequency is lower. 
Though filtering out these modes by setting the corresponding inverse power spectrum weights to zero, could avoid this issue, this procedure has to be consistently included in the calculation of the weight matrix, $(\mathbf{P}^{\,t} \mathbf{N}^{-1} \mathbf{P})^{-1}$, if an unbiased map estimate is to be derived. This, in turn, entails an effective increase of the actual bandwidth of the inverse noise correlation matrix, $\mathbf{N}^{-1}$, and typically enhances the level of noise correlations of the produced map. 
The trade-off in this case is therefore among the magnitude of the bias, the noise matrix bandwidth, and the complexity of the pixel-domain noise.

If the actual noise spectrum is more complex, and in particular, if it has some narrow features, then these also are affected as a result of imposing the band-diagonality.

\bibliographystyle{latex_style/aa}
\bibliography{bib_file/midas_prec}

\end{document}